\documentclass{aa}  

\usepackage{graphicx}
\usepackage{txfonts}
\usepackage{natbib}
\usepackage[colorlinks=true,     linkcolor=blue, citecolor=blue, filecolor=blue, urlcolor=blue]{hyperref}
\bibpunct{(}{)}{;}{a}{}{,} 
\newcommand{\DC}{DC303.8}
\newcommand{\DCL}{DC303.8-14.2}
\newcommand{\caii}{Ca\thinspace {\small II}}
\newcommand{\caiitr}{Ca\thinspace {\small II} triplet}
\newcommand{\onoff}{\em on--off}
\defcitealias{mattilaetal2018}{mat18}
\defcitealias{lehtinen_mattila_2013}{Leh2013}
\defcitealias{gordon_2004}{gor04}
\newcommand{\cgs}{$10^{-9}$erg cm$^{-2}$s$^{-1}\AA^{-1}$sterad$^{-1}$} 
\newcommand{\nW}{nW m$^{-2}$sterad$^{-1}$} 
 
\begin{document}


\title{Measurement of the extragalactic background light at 8600\thinspace \AA\,\\
   using dark cloud shadow and the Ca\,{\Large II}-triplet lines\thanks{Based on observations done at the European Southern
       Observatory, La Silla Paranal Observatory, Chile (programs 099.A-0028, 0102.A-0280 and 0104.A-0192)}} 

 \titlerunning{EBL at 8600 \thinspace \AA\,}

   \author{L.K. Haikala
          \inst{1}
          \and
          K. Mattila
          \inst{2}
          \and
          P. V\"ais\"anen
          \inst{3,4}
          }

   \institute{Instituto de Astronomia y Ciencias Planetarias,
          Universida de Atacama,
          Copayapu 485, Copiapo, Chile\\
              \email{lauri.haikala@uda.cl}
         \and
         Department of Physics, University of Helsinki, Gustaf
         H\"allstr\"ominkatu 2, FI-00014 Helsinki, Finland\\
             \email{kalevi.mattila@helsinki.fi}
         \and
         Finnish Centre for Astronomy with ESO, FINCA, FI-20014 University of
         Turku, Finland
         \and
         South African Astronomical Observatory, P.O. Box 9, Observatory, 7935,
         Cape Town, South Africa }

 
  \abstract{ We present the results of a measurement of the near-infrared
    extragalactic background light (EBL).  The surface brightness
    towards  the opaque intermediate-latitude globule DC303.8-14.2 was
    obtained using ESO {VLT}/FORS spectrophotometry.  Long-slit
    spectra covering the opaque core and the almost unobscured area
    north of the cloud were measured using the 
      nodding-along-the-slit  measuring technique, thus providing a
    differential spectrum  opaque core -- transparent area.  It
    is free of most of the foreground components, also excluding  most
    of the airglow time variations.  The scattered integrated
    starlight (ISL) from the dark core itself is the only remaining
    major foreground component when extracting the EBL from the
    differential spectrum.  The scattered starlight spectrum in the
    wavelength domain 8450 -- 8700\thinspace \AA\, is dominated by the
    strong \caiitr\ Fraunhofer lines at $\lambda$ 8498, 8542,
    8664\thinspace \AA\,, whereas the integrated light of galaxies
    and other contributors to the EBL intensity produce a smooth
    spectrum without these lines. We used the GAIA RVS spectral
    database to construct a template for the scattered ISL spectrum;
    another template was obtained by using the globule's
    semi-transparent bright rim.  The resulting EBL intensity as
    derived from the $\lambda$ 8542\thinspace \AA\ line is $I_{\rm
      EBL}= 1.62\pm 0.76(\sigma_{stat})$\,\cgs\, or
    $13.8\pm6.5(\sigma_{stat})$\,nW m$^{-2}$sr$^{-1}$; this represents
    a tentative detection at $2.1\sigma$\, level; the scaling
    uncertainty is $\pm 10\%$. }

   \keywords{Cosmology: diffuse radiation -- Galaxy: solar neighbourhood, stellar content
 -- ISM: dust, extinction -- Methods: observational --  Techniques: spectroscopy}

   \maketitle
%
\nolinenumbers
\section{Introduction}
\label{sect:intro}

The UV/optical/near-IR extragalactic background light (EBL) contains a
large fraction of the energy released in the Universe since the
re-ionisation epoch. The importance of the EBL is emphasised by the
detection of the cosmic far-infrared background, the dust-processed
fraction of the EBL \citep{hauseretal_1998}. EBL puts important
constraints on the formation and early evolution of galaxies and the
star formation history of the Universe. The integrated galaxy light
{(IGL) puts a firm lower limit to the EBL and it has been determined
  with increasing accuracy in the last ten years by
  \citet{driveretal2016}, \citet{koushan21} and \citet{tompkins2025};
  the extrapolated part beyond the current $m_{\rm lim}$$\sim$$30$ mag
  may still remain somewhat uncertain. Reaching beyond the magnitude
  limit of the IGL, the EBL represents an inventory of all the light,
  especially the diffuse light, produced by nucleosynthesis in stars
  outside the galaxies plus accretion into active galactic nuclei
  (AGN).} Low surface brightness galaxies and individual intergalactic
or interhalo stars are possible contributors to the
EBL. Moreover, the hypothetical decaying of elementary
  particles in a suitable energy range has been discussed in the
  past.

The measurement of the optical EBL has turned out to be a tedious
problem because the foreground components, the zodiacal light (ZL) and
airglow (AGL), are much brighter than the EBL.  We have recently
\citep{mattila_vaisanen_etal_2017} achieved a measurement of the EBL
at 400\,nm, utilising the shadowing effect of a dark nebula on the
background light.  A differential measurement of the surface
brightness of a high-latitude dark nebula and its surrounding area,
which is (almost) free of obscuring and scattering dust, provides a
signal that is free of ZL and AGL and is due to two components only:
(1) the EBL and (2) the integrated starlight (ISL) diffusely scattered
from interstellar dust in the cloud and, to smaller degree, from its
surroundings.

\begin{figure}
   \centering
\vspace{-5mm}
\hspace{-10mm}
   \includegraphics[width=8.0cm, angle=-90]{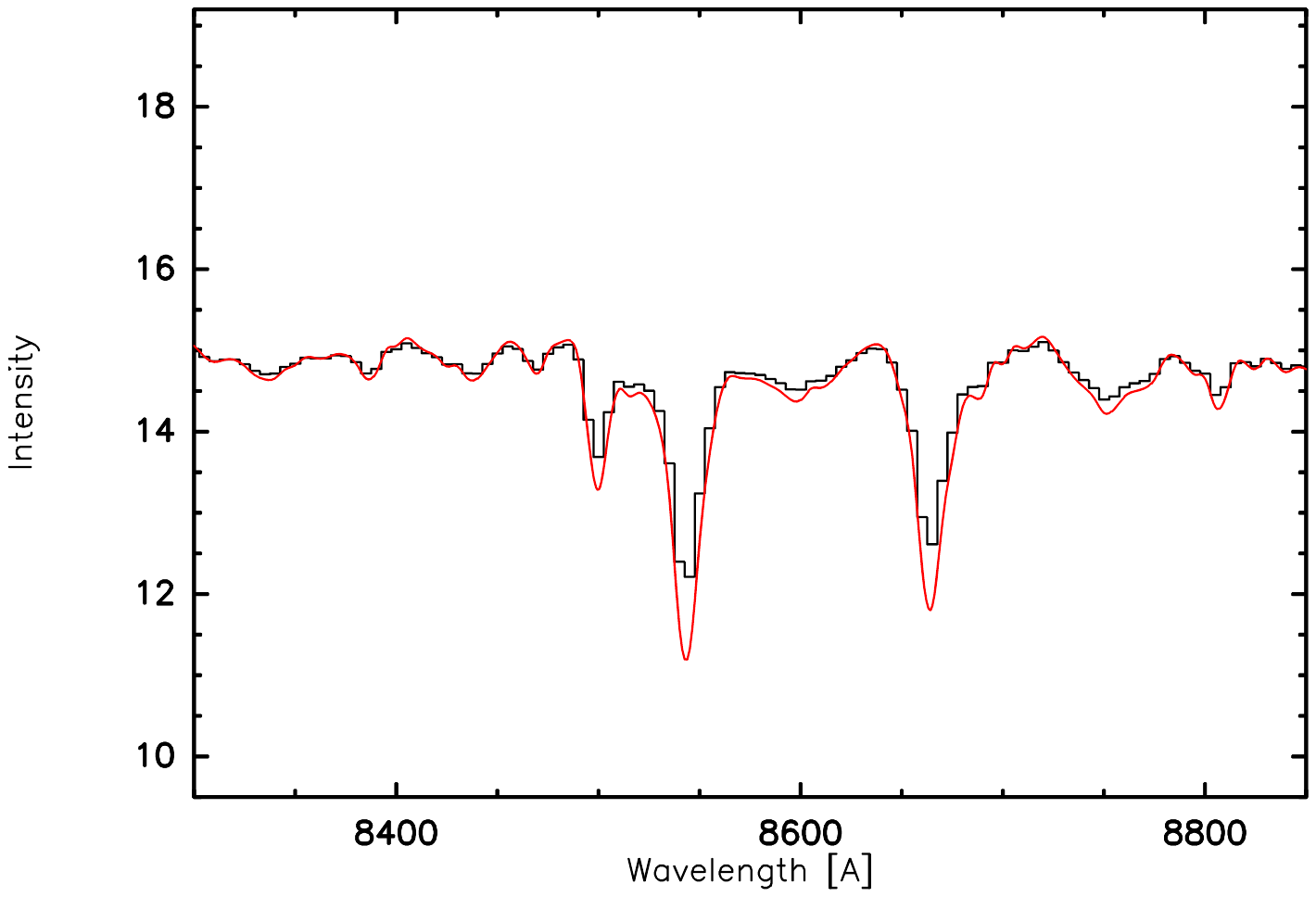}
\vspace{-10mm}
      \caption{
Model for the \caiitr\ lines in the synthetic scattered integrated starlight
(ISL) spectrum (black line) for an assumed continuum intensity of 15 units 
and a spectral resolution of 10\thinspace \AA\,; the Pickles spectrum library has been adopted  
\citep{lehtinen_mattila_2013}. The dashed red line shows the {\em on--off} spectrum (opaque dark 
core minus transparent background sky) in the presence of an EBL of 5 units.
              }
         \label{eblmodel}
   \end{figure}

The key issue of the dark cloud method is using the characteristic
absorption line spectrum of the scattered light which, for the line
depths, is a copy of the ISL spectrum. The EBL and ISL have different
spectra: the ISL spectrum has the characteristic Fraunhofer lines and
discontinuities (such as the Balmer and 400\,nm jumps), while the EBL
spectrum is  smooth; summing up radiation over a large redshift
range washes out the spectral features. The transmitted EBL signal is
present in the transparent OFF area only. It has the effect that in
the {\em on--off}  spectrum (opaque core minus transparent area) the
Fraunhofer line depths, measured in units of the adjacent continuum
level of the spectrum, are larger than they are in the pure scattered
ISL spectrum (see Fig. \ref{eblmodel}).

The tentative EBL detection at 400\thinspace nm
\citep{mattila_vaisanen_etal_2017} has suggested that there remains
space for an unknown EBL component of approximately equal magnitude to
the well-established IGL. An excess EBL component at $0.8 -
1.6\thinspace\mu m$, approximately equal to the IGL, has been
announced by \citet{matsuuraetal2017}. More recently,
{\citet{zemcov2017}, \citet{laueretal2021, laueretal2022},
  \citet{symonsetal2023}, and \citet{postman24}, using the {\em
    New-Horizons}-LORRI  broad-band imager 440 -- 870\,nm, have found
  different EBL values, varying from approximately two times the IGL down to no
  excess above the IGL level.}

\begin{figure*}
\sidecaption
  \includegraphics[width=12cm]{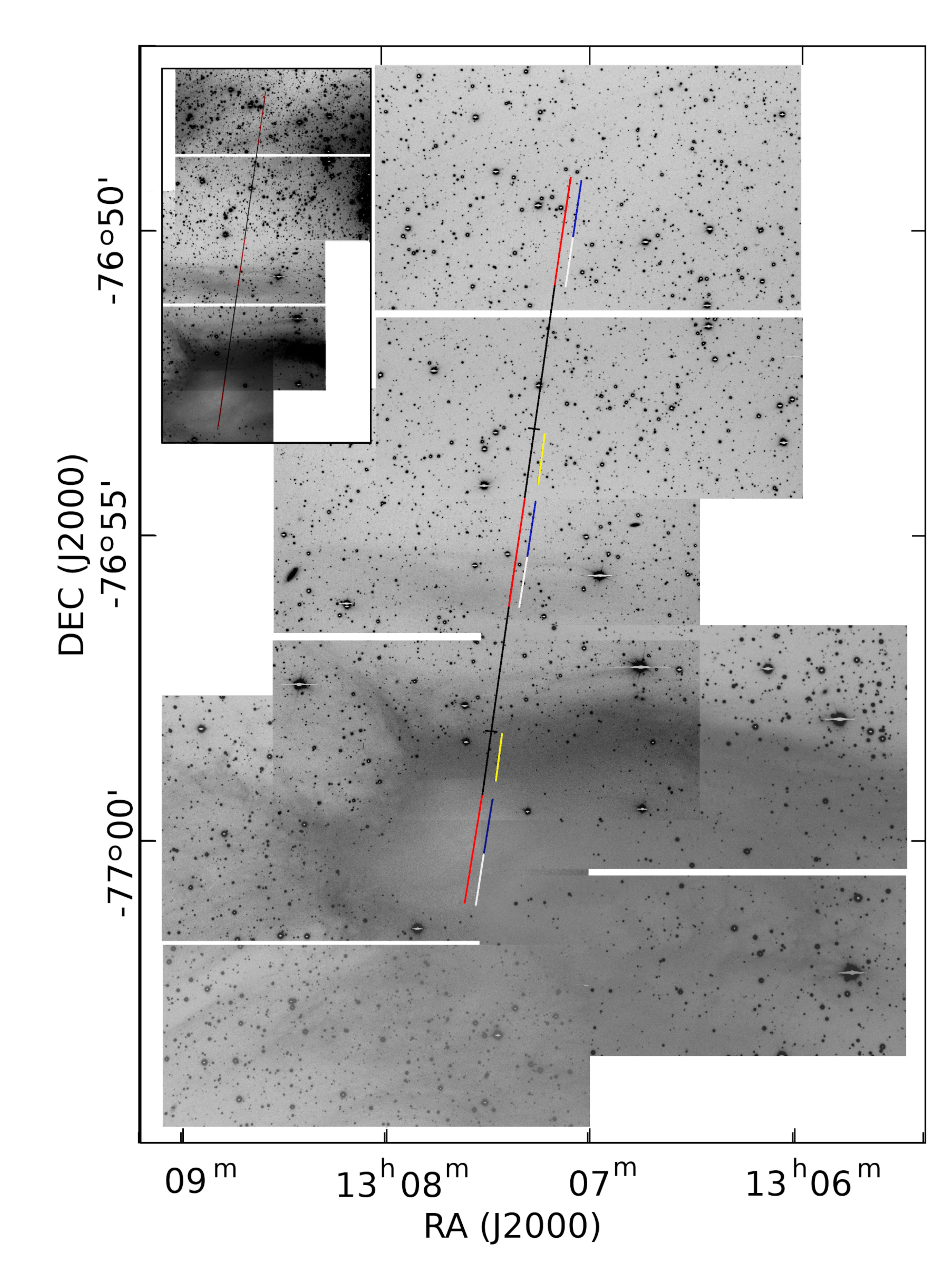}
      \caption{Mosaic of the I--BESS images observed during the
        \DC\ pre-imaging.  The position of the two {overlapping} FORS
        longslit positions on the sky during the nodding are
        superposed on the image. The overlapping 100\arcsec\ sections
        are indicated in red. The southernmost red section covering
        the globule core marks the position of the southern
        100\arcsec\ of the FORS2 CHIP2 during the first nod
        integration.  The northernmost red section marks the northern
        100\arcsec\ part of CHIP1 during the second nod
        integration. The centermost red section is covered first by
        CHIP1 during the first nod integration and by the CHIP2
        100\arcsec\ section during the second nod integration. {The
          slit sections discussed in the text are indicated in white,
          blue, and yellow.}}
   \label{figure:slits}
 \end{figure*}

In order to further constrain this unknown EBL component and also to
provide independent support for EBL detections{, ours and 
the others}, we need a spectral coverage that is as wide as
possible.  Because of the continuum and cumulative Lyman line
absorption by integalactic H{\small I} clouds, the EBL at 400\,nm
measures the EBL only up to the redshift of z $\sim$ 3. At 860\,nm, the
range is strongly expanded, up to z $\sim7.5$.  The aim of the present
study is to apply the dark cloud shadow method to longer wavelengths.

The strong \caiitr\ $\lambda$\,8498/8542/8662 offers the best
possibility in the red--near-IR wavelength range 550 – 1000\,nm. {Along
  a similar line of thought, \citet{korngut22} have used the
  \caiitr\ line 8542\thinspace \AA\, to measure the absolute intensity
  of the ZL, a serious foreground hindrance for the measurement of the
  EBL.}

The high latitude globule \object{DC303.8-14.2} \citep {DC1986}, first
listed as \object{Sandqvist 160} \citep {sandqvist1977}, located at
$\alpha$ = 13:07:40.0, $\delta$ = -77:00:00 (J2000), is a suitable
target for the application of the dark cloud method.  It has a large
core depth, $\tau(860$\,nm\,$) \ge 10$, and OFF areas with sufficient
transparency, $\tau(860$\,nm$) \le 0.5$, are available outside the core
\citep{kainulainenetal2007}.  Low-resolution spectra have already been
obtained with NTT+EFOSC by \citet{lehtinen_mattila_2013}.

\section{Observations and data reduction}
\label {sect:Obs}

The FORS2 \citep{appenzeller98} \DCL\ pre-imaging and the LSS
spectroscopic data were acquired in service mode during 2017, 2019,
2020, and 2021 on the VLT UT1 telescope Antu. The instrumental set-up
of the spectroscopy is listed in Table \ref{setup}.  Due to the
faintness of the EBL signal and the brightness of the expected {sky
  background, dark time and photometric sky conditions were
  requested. Because of the southern declination of the target (-77
  deg), the observations were limited to time slots of $\la6$ h.
  Standard stars were observed during the observations to calibrate
  the spectrometer response.

\subsection { The observation strategy}
\label{strategy}

\DCL\ and the background to the north was pre-imaged with FORS2
(I\_BESS filter) in order to find an optimal position for the
longslit{, i.e.} avoiding stars as much as possible (see
Fig. \ref{figure:slits}).  Due to the strong {and variable} airglow
spectral lines in the 800 -- 900\,nm, region and high background sky
brightness of $\gtrsim$200 -- 300 {\cgs \thinspace (hereafter
  cgs-unit) the measurement of the expected $I_{\rm
    EBL}$ signal of  1 -- 5 cgs-units is technically very demanding;
  for example, it requires an exceptionally good flat-fielding
  accuracy of $\sim0.2\%$, when applying the simple slit technique.
  However, such high accuracy cannot be reached with standard
  flat-field procedures.  Therefore, instead of a simple single
  long-slit spectrum across the nebula, the following more
  sophisticated strategy was used. The core of the globule and the
  adjacent transparent comparison areas were observed using the same
  part of the slit (and {exactly} the same detector pixels). This was
  achieved by nodding the telescope relatively rapidly back and forth
  by 300\arcsec\ along the 6\farcm8 (408\arcsec) long slit so that the
  slit {positions} overlapped by 100\arcsec\ during individual nods
  (Fig. \ref{figure:slits}). The night-sky brightness at 850 -- 864\,nm
  is dominated by airglow which varies on timescales of minutes. An
  {\em on--off} cycle time of 5 minutes can be done with 90 seconds
  {\em on}, 90 seconds {\em off}, when nodding is done along the slit
  (ESO observing template FORS2-lss\_obs\_off\_fast).  The
  \DCL\ spectroscopy data were collected during 27 observing blocks of
  11 nods each.  The number of observations in each year and month are
  listed in Table \ref {Table:nblocks}.

\subsection {Data reduction}
\label{reduction}

The ESO esoreflex-2.9.1 pipeline and the ESO Common Pipeline library,
EsoRex, were used to reduce individual spectra. The pipeline was used
to analyse the standard star measurements and to produce the FORS2
efficiency and response tables during the observing runs. EsoRex
fors\_calib was used to rectify and wavelength calibrate the spectra
and to produce the MASTER\_BIAS, the MASTER\_NORM\_FLAT\_LSS images,
and the GLOBAL\_DISTORTION\_TABLE.  Flat-fielding must be applied if
the {\em on--off} position continuum level difference is not zero even
if the same detector pixels are used. The MASTER\_NORM\_FLAT\_LSS was
smoothed to remove the high frequency noise, but conserving the
flat-field large-scale gradients. The target spectra were first bias
subtracted, then flat-fielded using the smoothed
MASTER\_NORM\_FLAT\_LSS and remapped using EsoRex fors\_science, but
not flux calibrated

The MAPPED\_ALL\_SCI\_LSS spectra provided by fors\_science were
collected for each observing block.  These spectra still have the
strong sky airglow lines. The sky subtraction was done by subtracting
the average of the two {\em off} spectra taken at 300\arcsec\ north,
one immediately before and one after the {\em on} spectrum. This was
done separately for FORS2 detectors, Chip1 (A), and Chip2 (B). Thus,
each {of the 27} observing blocks of 11 telescope nods provided ten
{\em on--off} spectra for the two detector chips.  The Chip2 {\em
  on--off} spectra provide the difference between the globule core and
bright rim areas and the
middle {\em off} area positions, while the Chip1 {\em on--off} spectra
provide the difference between the middle {\em off} area and the
northernmost {\em off} area positions (for the location of the areas,
see Fig. \ref{figure:slits}).  {After the {\em on--off} subtraction
  the flat-fielding affects only the difference spectrum as the bulk
  of the atmospheric continuum and AGL intensities cancel out.}  The
{\em on--off} spectra from the 27 observing blocks were first cropped
so that the pixels in the spectra aligned correctly and then collected
to a single database.  The final database contained 270 reduced {\em
  on--off} spectra for CHIP1 and CHIP2 each.

The ADU difference spectra were corrected for aperture efficiency and
flux calibrated using the spectral response tables from standard star
measurement nearest in time. {These spectra, now in units of erg
  cm$^{-2}$s$^{-1}\AA^{-1}$ per readout pixel (binned by 2), were then
  converted to surface brightnesses. The solid angle of a pixel is
  given by slit width ($2.5\arcsec$) $\times$ pixel size ($0\farcs
  25$) = 1.469 $10^{-11}$ sterad.  }

The 270 {\em on--off} spectra were median averaged for the A and B
detectors. Using the median rather than the average helped to
eliminate extreme AGL time variability outliers and cosmic-ray hits.
The resulting averaged 2D spectra contain numerous stars, most of
which are not visible in the original single {\em on--off}
spectra. Because of the high extinction in the direction of \DCL\ the
number of visible stars in the B detector averaged spectra is lower
than that in the OFF region covered by detector A. The stars and the
remaining instrumental artefacts were cleaned from the averaged
spectra.  Finally 1D spectra covering the slit sections discussed in
the text (see Table \ref{Table:slitpos} and Fig. \ref{figure:slits})
were formed by averaging the 2D median averaged spectra in the slit
direction using the IRAF routine blkavg.  The average
was adopted because the time-variability aspect does not occur in the
median averaged 2D spectra and because of the somewhat smaller rms
compared to the median. In the used instrumental set-up, the FORS
spectral resolution is 9.0\thinspace \AA, but the readout dispersion
(after binning by two) is 0.84\thinspace \AA. Thus, the 1D spectrum
was block averaged by four pixels in the dispersion direction.  The
resulting surface brightnesses of the Core-up and Bright\_rim 
sections were then $\sim$19 and $\sim$37\thinspace cgs-units,
respectively. These values are in reasonably good agreement with the
surface brightnesses observed with EMMI at the New Technology
Telescope at La Silla \citep{lehtinen_mattila_2013}.

\begin{figure}
\centering
\includegraphics[width=90mm]{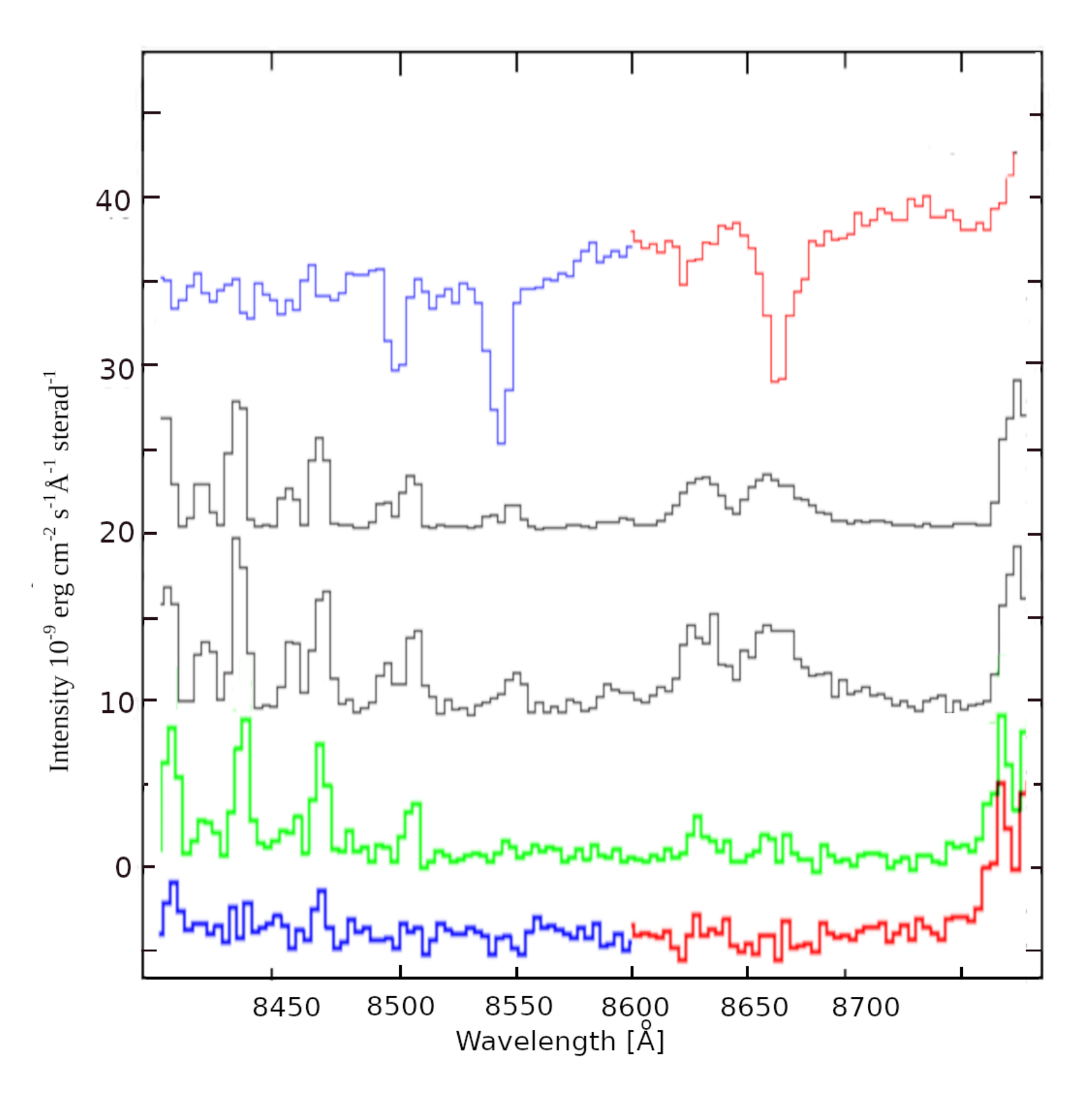}
\caption{ From top to bottom: Chip 2 extracted Bright\_rim spectrum
  combined from 179 {\em on--off} spectra below 8600\thinspace \AA
  \ (in blue) and 228 {\em on--off} spectra above 8600\thinspace
  \AA\ (in red).  A typical airglow spectrum scaled by 1/500 and
  shifted upwards by 20\thinspace cgs-units. The standard deviation of
  the intensities along the slit in the Chip 2 Bright\_rim 2D spectrum
  shifted upwards by 4\thinspace cgs-units. The
  A\_100\arcsec-off\ spectrum (green) from the sum of all 270 {\em
    off--on--off} spectra; it has been shifted upwards by 3\thinspace
  cgs-units. The A\_100\arcsec-off\ spectra for the sub-sets of 179
  and 228 individual spectra, selected for minimum-AGL residuals in
  blue and red for $\lambda < 8600$\thinspace \AA\, and $\lambda >
  8600$\thinspace \AA\,, respectively.}
\label{figure:noise}
\end{figure}

 The primary source of noise in the 1D {\em on--off} spectra is due to
 atmospheric emission and the airglow lines during nodding. Different
 from the atmospheric continuum, the emission due to airglow lines
 varies randomly during the nodding producing 1/f noise.  If the
 variation of the airglow lines is linear in time, using the average
 of the two off spectra taken immediately before and after the on
 spectrum as the off spectrum helps to suppress the airglow noise in
 the {\onoff} spectrum.  The \caiitr\ lines coincide in wavelength
 with strong airglow lines, the 8498\thinspace \AA \ and
 8542\thinspace \AA \ lines {(hereafter lines 1 and 2, respectively)}
 with the hydroxyl (OH) molecule ro-vibrational 6--2 transition lines,
 and the 8662\thinspace \AA \ line { (hereafter line 3) }with broad O2
 (1--0) lines.  The intensity of residuals due to the O2 line were
 generally lower than those of the OH lines; in addition, the OH and
 O2 airglow line residuals varied at times in opposite directions.

A visual selection of the least contaminated {\em on--off} spectra was
made separately for the wavelength ranges 8500 -- 8600\thinspace \AA
\ and 8600 -- 8700\thinspace \AA, which overlap lines 1 + 2 and line
3, respectively. The first sub-set contains 179 and the second
contains 228 spectra.  The relation between the airglow lines and the
noise in the summed spectra is demonstrated in Fig. \ref
{figure:noise}. Although the strengths of the AGL bands in the {\em
  on--off} spectra have been reduced (by a factor of $\sim 10^{-2}$)
compared to their total sky levels, the spectra are still contaminated
by the AGL residuals at a low level. Elimination of the spectra most
strongly affected by airglow variations strongly suppresses the {AGL
  residual} noise in the low extinction Chip 1 summed up
A\_100\arcsec-off position where no \caiitr\ lines are expected
(Fig. \ref {figure:noise}).  All the {final} spectra in the slit
sections discussed in this paper are shown in Fig. \ref{figure:all}

   \begin{table}
      \caption[]{The FORS2 instrumental set-up.}
         \label{setup}
\begin{tabular}{ll}

           \hline
           \hline
            Telescope/Instrument& VLT-UT1/FORS2                                       \\
            mode                    & LSS                                             \\
            Grism                   & GRIS\_1028z                                     \\
            Filter                  & OG590                                           \\
            Spectral resolution     & 9.0\thinspace \AA\,FWHM                         \\
            Detector                & 2$\times$ MIT                                   \\
            Size                    & 2$\times$ (2048$\times$  1024) \tablefootmark{a}\\
            LSS slit width          & 2\farcs5                                        \\
            LSS slit length         & 6\farcm8                                        \\
            LSS position angle       & 8\degr                                         \\
            Spatial scale           & 0\farcs25/pixel \tablefootmark{a}                \\
            Dispersion              & 0.84\thinspace \AA/pixel\tablefootmark{a}       \\
            Wavelength range        & 7900\thinspace \AA - 9500\thinspace \AA                               \\ 

            \hline
\tablefoottext{a}{After binning the CCDs by 2$\times$2.}
\end{tabular}
\end{table}

\begin{table}
\vspace{1cm}
      \caption[]{Names used for the {\em on-off} spectra taken along the long-slit shown in Fig. \ref{figure:slits} and their extensions.
}
\begin{tabular}{lllll}
  \hline
  \hline
  Name             & Pos. {\em on}\tablefootmark{a}   & Pos. {\em off\ } \tablefootmark{a} &   Colour    & Cleaned  \tablefootmark{a}   \\
                 &               &              &     in Fig.2  &  width   \tablefootmark{b}         \\

\hline 
Core-100\arcsec    &   0\arcsec - 100\arcsec      & 300\arcsec-400\arcsec  &      red              &    86''        \\
Core-low           &   0\arcsec -50\arcsec        & 300\arcsec-350\arcsec  &      white            &    46''        \\
Core-up            &  50\arcsec - 100\arcsec      & 350\arcsec-400\arcsec  &      blue             &    40''        \\
Bright\_rim        & 115\arcsec - 161\arcsec      & 415\arcsec-461\arcsec  &      yellow           &    38''        \\
A\_100\arcsec-off  & 300\arcsec-400\arcsec        & 600\arcsec-700\arcsec  &      red              &    64''        \\
A\_low-off         & 300\arcsec-350\arcsec        & 600\arcsec-650\arcsec  &      white            &    31''        \\
A\_core-off        & 350\arcsec-400\arcsec        & 650\arcsec-700\arcsec  &      blue             &    33''        \\
\hline                          
\end{tabular}
\tablefoottext{a}{The {\em on} and {\em off} coordinates are counted from the southern end of the CHIP2 slit covering the globule core.\newline}
\tablefoottext{b}{Slit section width after cleaning stars}

\label {Table:slitpos}
   \end{table}

\section{Separation of the EBL and the scattered light}

The dust in DC303-14.2 acts as an obscuring screen in front of the
extragalactic background light.  The attenuation of the EBL depends on
the line-of-sight extinction: blocking is strong towards the opaque
core, while less shadowing occurs in the transparent outer areas.  The
scattered starlight dominates over the effects of the EBL in the
observed {\em on--off} surface brightness difference.  The situation
in the optical differs from that with the X-ray shadows seen towards
dark clouds and globules { \citep [see e.g.][] {freyberg2004,
    Yeung2023}} where the attenuation is caused by pure absorption of
cold gas in the cloud.

In order to disentangle the EBL and the scattered starlight components
one can make use of spectroscopic observations in suitable wavelength
slots where the scattered starlight has strong spectral features,
absorption lines or discontinuities.  For the present study we used
the \caii\ absorption line triplet at $\lambda$8499/8542/8662. It is
one of the strongest spectral features in the red/near-IR integrated
starlight, and is therefore well suited also for the study of low
surface brightness targets.  The triplet is prominently present in our
scattered light spectra of DC303.8-14.2.

\subsection{The components of the observed {\em on--off} spectrum}
\label {sec:components}}
The observed {\em on--off} surface brightness difference, $\Delta
I_{\rm obs}(\lambda)$, has two components: the scattered starlight,
$I_{\rm sca}(\lambda)$, and the extragalactic background light,
$I_{\rm EBL}$. The observed difference can be expressed as
\begin{eqnarray}
\lefteqn{{\Delta I_{\rm obs}(\lambda)} = {I^{\rm on}_{\rm sca}(\lambda) - I^{\rm off}_{\rm sca}(\lambda) 
+ I_{\rm EBL}(e^{-\tau}+f_{\rm sca}(\tau))} - I_{\rm EBL}
 ={} } \nonumber \\   
  & &  {}= {\Delta I_{\rm sca}(\lambda)-I_{\rm EBL}h(\tau)}
\label{Eq1},
\end{eqnarray}
where $I^{\rm on}_{\rm sca}(\lambda)$ and $I^{\rm off}_{\rm
  sca}(\lambda)$ are respectively the intensities of starlight as
scattered by dust in the cloud and in the semi-transparent {\em off}
area; the term $I_{\rm EBL}e^{-\tau}$ stands for the transmitted EBL;
and $I_{\rm EBL}$ is assumed to be constant over the wavelength range
relevant in this study $\lambda = 8450 - 8700$\thinspace \AA\,.

Not only the starlight, but also the photons of the isotropic EBL are
scattered by the dust; this scattered EBL component is given by the
term $I_{\rm EBL} f_{\rm sca}(\tau)$.  Here, we introduce the notation
$h(\tau) = 1-e^{-\tau}-f_{\rm sca}(\tau)$, which we call the
attenuation factor. It gives the fraction by which the EBL is
attenuated in the direction of the cloud relative to a transparent
{\em off} area: $h(\tau) = 1$ for the case of complete obscuration and
$h(\tau)= 0$ for no obscuration.  For small and moderate optical
depths, $\tau \la 1-2$, scattering compensates much of the cloud's
obscuring effect. Even for high opacities it still amounts to a
substantial fraction of $I_{\rm EBL}$.  For the high-opacity core area
of DC303-14.2, the scattered light seen towards the core is dominated
by the outer layers up to $\tau \approx 2 - 3$~mag, whereas the core
itself dominates the attenuation term $e^{-\tau}$.

Values of $f_{sca}(\tau)$ for different cloud opacities and dust
scattering properties have been calculated using Monte Carlo radiative
transfer modelling \citep {mattila1976}.  However, such results depend
strongly on the adopted dust-scattering parameters. Therefore, we have
estimated the $f_{sca}(\tau)$ values empirically, by using the
scattered starlight values, $I_{\rm sca}(\lambda)$, for the
Core-100\arcsec\ and the Bright\_rim of \DCL\, as a guide. Based on
these estimates we have derived for the Core-100\arcsec\ and
Bright\_rim areas the values { $h(\tau)$ = 0.92 and 0.69,
  respectively. Their systematic errors were estimated to be 2\% and
  4\%, respectively.  For details we refer to Appendix
  \ref{App:shadowing}.}

\subsection{Model fitting of the opaque core spectrum} 

The observed spectrum $\Delta I_{\rm obs}(\lambda)$ is fitted in
accordance with Eq. \ref {Eq1}.  The spectrum of the scattered starlight
from the cloud can be represented as the product of two factors,
\begin{equation}
I^{\rm on}_{\rm sca}(\lambda)= i_{\rm ISL}(\lambda) G^{\rm on}_{\rm sca}(\lambda),
\label{Eq2}
\end{equation}
where $i_{\rm ISL}(\lambda)$ stands for the normalised spectrum of the
impinging integrated starlight. It is normalised to 
1 at a reference wavelength $\lambda_0$:
\begin{equation}
 i_{\rm ISL}(\lambda) =  I_{\rm ISL}(\lambda) / I_{\rm ISL}(\lambda_0).
\label {Eq3}
\end{equation}

Here $G^{\rm on}_{\rm sca}(\lambda)$ accounts for the intensity and
gradient (reddening or bluing) of the scattered light relative to the starlight
spectrum, $i_{\rm ISL}(\lambda)$, as caused by the
wavelength-dependent scattering and extinction in the cloud.  $G^{\rm
  on}_{\rm sca}(\lambda)$, is assumed to be, over a
  limited wavelength range, linear function of $\lambda$.

\begin{equation}
  G^{\rm on}_{\rm sca}(\lambda) = G^{\rm on}_{\rm sca}(\lambda_0)[1+grad^{\rm on}\times(\lambda-\lambda_0)
\end{equation}
We write in a similar way for the {\em off}\, position,

\begin{equation}
G^{\rm off}_{\rm sca}(\lambda) = G^{\rm off}_{\rm sca}(\lambda_0)[1+grad^{\rm off}\times(\lambda-\lambda_0)
\end{equation}
and for the difference {\em on--off}\, 
\begin{equation}
\Delta G^{\rm on-off}_{\rm sca}(\lambda)= \Delta G^{\rm on -off}_{\rm sca}(\lambda_0)[1 + grad(\lambda-\lambda_0)
\end{equation}
 where $grad =  grad^{{\rm on}-{\rm off}}$.}
 
For the fitting of the observed {\em on--off}\, surface brightness
difference, $\Delta I_{\rm obs}(\lambda)$, as given by Eq. \ref {Eq1},
we have used the {\sc
  idl}
programme {\tt
  MPFITFUN}.\footnote{https://cow.physics.wisc.edu/\texttt{\char`\~}craigm/idl/mpfittut.html}
For the fitting it is represented as
\begin{equation}
  \Delta I_{\rm obs}(\lambda)= [p_0+p_1(\lambda-\lambda_0)]i_{\rm ISL}(\lambda) - h(\tau)I_{\rm EBL} .
\label{Eq7}
\end{equation}
Here the parameters $p_0$ and $p_1$ correspond to $G_{\rm sca}(\lambda_0)$ and $grad$.
For $\lambda_0$ we adopt 860.0\,nm. 

For the estimation of $I_{\rm EBL}$\ we adopted two different
approaches: First, we constructed the integrated starlight spectrum
$i_{\rm ISL}(\lambda)$ by adding up a large number of stellar RVS
spectra as provided in the GAIA3 Data release (see Section
\ref{sec:model_fit}).

Secondly, we utilised the observed spectrum of the bright rim. Because
of the different values of the attenuation factor $h(\tau)$, the core
and the rim spectra are influenced by different amounts of EBL; by
combining the two spectra we can determine $I_{\rm EBL}$ without
explicitly knowing the integrated starlight spectrum $i_{\rm
  ISL}(\lambda)$ (see Section \ref{sec:model_fit2}).

We note that while the scattered light has a continuum gradient
(reddening or bluing), which differs from the impinging ISL spectrum,
the depths, and profiles of the Fraunhofer lines, including the
\caiitr , remain unchanged.

\subsubsection{Model fit using GAIA RVS integrated starlight spectrum}
\label {sec:model_fit}
The interstellar radiation field (ISRF) in the optical and
near-infrared is mainly produced by the integrated starlight (ISL),
the sum of light from individual stars distributed over all directions
and magnitudes. A modest (10 -- 20\,\%) indirect contribution is added
by the diffuse galactic light (DGL), that is starlight scattered
off interstellar dust.  The absorption line spectrum of the DGL is the
same as that in the ISL.

The recent Gaia Data Release 3 has made available the Radial Velocity
Spectrometer (RVS) mean spectra for $\sim$1 million stars brighter
than $G_{RVS} \sim$14\,mag ($G \sim $15mag), distributed all over the
sky.  We have derived the integrated starlight spectrum for a number
of selected areas by adding up the contributions of all stars with
$G_{RVS} \le $12mag, which is the limit of completeness for the mean
RVS spectra in Gaia DR3.

Circular areas were selected with diameters 8\degr\ -- 20\degr,
depending on the star density in the area. In view of the position of
our target globule at $l = 303\fdg 8, b = -14\fdg 2$, areas were
chosen at latitudes $0, \pm10, -15, \pm30, \pm60$\degr\ for each of
the longitudes 240, 270, 300, 330, 360\degr, plus the polar caps, $b =
\pm90$\degr.  The number of stars was mostly 4000 to 5000 per
area. The selection of these GAIA RVS areas was done with the
scattering properties of the interstellar dust in mind.

Observations of the DGL and scattered light in dark nebulae have shown
that the scattering function $S(\Theta)$ is strongly
forward-throwing. It is frequently approximated by the analytic
expression according to \citet{henyey}, characterised by the asymmetry
parameter $g = \langle {\rm cos}\Theta \rangle$.  Observations have
indicated that $g \approx 0.6 - 0.9$ \citep [see review by][]
{gordon_2004}. Specifically for globules and dense cloud cores the
observational results have been more restrictive, however.  The
asymmetry parameter values for them at optical to near-IR wavelengths
range between $g = 0.7$ and 0.9 \citep{mattila_1970, witt1974,
  witt_1976, witt_1990, togi_2017,mattilaetal2018}. We adopt for our
estimates for \DCL\ the value $g = 0.75$. In order to test the effect
of a less forward-throwing scattering function we have also calculated
the integrated starlight spectrum for another, albeit unrealistic,
case of $g = 0.5$.

For $g = 0.75$ the illumination of the globule comes from a relatively
narrow cone: 65\% comes from $\Theta < 30\degr$ and 77\% from $\Theta
< 40\degr$.  The distribution of our selected sky areas, as listed
above, covers this range well.  We divided the sky between
$240\degr<l<360\degr$ and $-90\degr<b<90\degr$ into 42
`parcels', each one with a GAIA RVS area in its centre.  For each GAIA
RVS area the fluxes, $j^i_k(\lambda)$, of all GAIA RVS stars with
$G_{RVS} \le 12$\,mag were added up, and the sum was normalised to 1
at $\lambda_0 = 860$\,nm. It is denoted by $\sum_i
j^i_k(\lambda)$. This normalised sum spectrum represents the spectrum
of the whole parcel.  To each parcel, k =1,42 a weight was assigned
according to 

\begin{equation}
W_k = A_k \times \bar{I}_k^{\rm ISL} \times S_k^{\rm H-G}(\Theta_k)
\end{equation}
    
where $A_k$ is the area and $\bar{I_k}^{\rm ISL}$ the mean ISL surface
brightness of the parcel; $S_k^{\rm H-G}(\Theta_k)$ is the intensity
of the scattered light according to the Henyey--Greenstein scattering
function for $g = 0.75$.  The scattering angle $\Theta_k$ is the angle
between the directions towards parcel $k$ and the globule;
$\bar{I}_k^{\rm ISL}$ was taken from the Pioneer 10/11 $R$ band
starlight mapping (see \citet{Gordon98} and the  web 
page
\footnote{https://www.stsci.edu/\texttt{\char`\~}kgordon/pioneer\_ipp/Pioneer\_10\_11\_IPP.html}.

\begin{equation}
\LARGE \sum_{k=1}^{42}  W_k \times \large \Sigma_i j_k^{i}(\lambda)
\end{equation}

The weighted sum spectrum, normalised to 1 at $\lambda_0 = 860$\,nm,
was convolved with the FORS instrumental profile and resampled then to
the same channel width as the observed FORS spectra (see Appendix
\ref{App:convolution}). This spectrum was used as the ISL spectrum
$i_{\rm ISL}(\lambda)$ in Eqs. \ref{Eq2} and \ref{Eq3}.

Results for the Bright\_rim and Core-up spectra, fitted according to
Eq. \ref {Eq7}  with the GAIA RVS spectrum for $g = 0.75$, are shown in Figure
\ref{Fig_4}.  In the lower part of Fig. \ref{figure:newgaia} we
compare the GAIA interstellar radiation field spectra for the two
cases $g = 0.75$ and $g = 0.50$. The depths of the three \caii\ lines
for $g = 0.50$ differ by no more than 0.5\% from the case $g = 0.75$.
It can be concluded that that even for the extreme g-value of 0.5 the
difference in the GAIA-based radiation fields for $g = 0.50$ and $g =
0.75$ would be much smaller than the observational uncertainties of
the line depths (see Fig. \ref{Fig_4}).  We conclude that the
uncertainty caused by the adopted g-value can be neglected.

Because a substantial fraction of the ISL comes from stars with
$G_{RVS} > 12$ mag, we have to address the question of how much light
is missing in our ISL spectra, and whether the RVS spectrum of the
stars with $G_{RVS} > 12$ mag differs from that of the brighter ones
with $G_{RVS} < 12$ mag.

We performed the star counts using the GAIA DR3 red magnitudes
$G_{RP}$.  The magnitude range covered was 2 -- 20\,mag The three
areas chosen for the counts, $(l,b)$ = (300,-5), (300,-15), and (300,
-30), covered the sky area most relevant for illumination of the
globule. The fraction of the ISL contributed by stars with $G_{RP} >
12$ is 48\%, 30\%, and 23\% of the total ISL for these three areas.

{Secondly, we constructed a synthetic model for the ISL spectrum.  The
  STELIB library was used for stellar spectra, and a simple Galaxy
  model was constructed with realistic values for the distribution
  parameters for stars of different spectral types and luminosity
  classes. The model and the results for $|b| = 11.5\degr$ are
  presented in Appendix \ref{App:ISLmodel}.

In summary, the results of the STELIB model show that the ISL spectrum
derived using the GAIA RVS spectra for the magnitude-limited sample,
$G_{RVS}\le 12$\,mag, can be considered as a good representation of
the total ISL spectrum as well.

\subsubsection{Model fit using the bright rim spectrum as template}
\label {sec:model_fit2}

From Eq. \ref{Eq7}, written for the bright rim spectrum, we can solve $i_{\rm ISL}$; we use it to replace
 $i_{\rm ISL}$ in the expression for the dark core:
\begin{eqnarray}
\lefteqn{\Delta I^{\rm DC}_{\rm obs}(\lambda)=  {} } \nonumber \\ 
& {}\frac{p_0^{\rm DC}+p_1^{\rm DC}\times (\lambda-\lambda_0)}{p^{\rm BR}_0+p^{\rm BR}_1\times (\lambda-\lambda_0)}\Delta I^{\rm BR}_{\rm obs}(\lambda)
     -[h^{\rm DC} - \frac{p^{\rm DC}_0+p^{\rm DC}_1\times (\lambda-\lambda_0)}{p^{\rm BR}_0+p^{\rm BR}_1\times (\lambda-\lambda_0)}h^{\rm BR}]\times I_{\rm EBL} 
\end{eqnarray}
Because the gradient terms $p_1^{\rm DC}\times(\lambda-\lambda_0)$ and 
$p_1^{\rm BR}\times(\lambda-\lambda_0)$ are much smaller than the terms $p_0^{\rm DC}$ and 
$p_0^{\rm BR}$, the expression of the dark core spectrum, to be used for the fitting with the 
bright rim spectrum, can to a good approximation be represented as
\begin{equation}
 \Delta I^{\rm DC}_{\rm obs}(\lambda)= [\frac{p_0^{\rm DC}}{p_0^{\rm BR}}+q_1 \times (\lambda-\lambda_0)]\Delta I^{\rm BR}(\lambda) - [h^{\rm DC}-\frac{p_0^{\rm DC}}{p_0^{\rm BR}}h^{\rm BR}]I_{\rm EBL} 
\end{equation}
The term $q_1\times(\lambda-\lambda_0)$ represents the weak wavelength
dependence of the dark core-to-bright rim ratio, as caused by dust
extinction and scattering in the globule.  For our observed spectra
the ratio ${p_0^{\rm DC}}/{p_0^{\rm BR}}$ is $\approx 0.49$. With the
actual attenuation factors of {$h^{\rm DC}\approx 0.92$} and $h^{\rm
  BR}\approx 0.69$ the resulting effective attenuation factor for the
dark-core fitting with the rim fitting becomes {$h_{\rm eff} = h^{\rm
    DC}-({p_0^{\rm DC}}/{p_0^{\rm BR}})h^{\rm BR} = 0.58$.} With the
absolute errors of $\pm 2\%$ and $\pm 4\%$ for $h^{\rm DC}$ and
$h^{\rm BR}$ the absolute (scaling) error for $h_{\rm eff}$ is $\pm
2\%$.

The GAIA-RVS fitting method (Section \ref{sec:model_fit}) has two advantages regarding the errors.
First, the integrated GAIA-RVS spectra have much smaller statistical errors than 
our  FORS bright-rim spectra; second, for the determination of $I_{\rm EBL}$ the larger value of 
$h(\tau)=0.92$ as compared to $h_{\rm eff}(\tau)=0.58$ results in smaller uncertainties for $I_{\rm EBL}$. 
However, the fitting of the dark core with the  bright-rim spectrum is less sensitive to systematic error sources.
The simultaneous acquisition of the dark core and bright rim spectra, within the same observing cycle and with
identical integration time slots, helps to suppress error sources caused by the different origins of the
observed and model spectra in the GAIA-ISL fitting method. The residual errors caused by the airglow 
time variations are the same in the dark core and bright rim spectra partially cancel out in the fitting
method.

\begin{table*}
\begin{minipage}{180mm}
 \caption{EBL values from {\tt MPFITFUN} fitting  utilising the three \caiitr\ lines of the scattered ISL: GAIA spectrum used as template in the fit}
\label {Table:results_GAIA}
\begin{tabular}{llllcl}

\hline
\hline
 Wavelength & Line(s)&FORS spectrum & rms   & $h(\tau)I_{EBL} \pm \sigma_{\rm stat}$ & $I_{EBL} \pm \sigma_{\rm stat}$ \tablefootmark{a}  \\
            &        &              & [cgs] &  [cgs]                             &      [cgs]                     \\
\hline

  8470 -- 8530      &    line 1  &   Core-up                &    0.72        &   4.69$\pm$4.32  &                        \\                                     
                 &            &   Core-low                &    0.62        &  -2.76$\pm$3.73  &                        \\                              
  &            &   Core-$100\arcsec$                    &    0.52        &   2.06$\pm$3.08  &                        \\
\hline
  8515 -- 8590    &    line 2  &   Core-up               &    0.48        &   0.55$\pm$1.29  &                        \\                              
                 &            &   Core-low                &    0.35        &   1.95$\pm$0.92  &                        \\                              
                 &            &   Mean of Core-up-low  &                &   1.48$\pm$0.75                  &   {1.61$\pm$0.82}        \\                                     
                 &            &    Core-100$\arcsec$                   &    0.26        &   1.50$\pm$0.70  &   {1.63$\pm$0.76}        \\                                     
                 &            &   Bright\_rim                          &    0.62        &  -0.31$\pm$1.69  &                        \\                              
\hline                                                                                                                 
  8630 -- 8690      &    line 3  &   Core-up                &    0.48        &  -0.52$\pm$1.59  &                        \\                              
                 &            &   Core-low                &    0.90        &   4.29$\pm$2.95  &                        \\                              
                 &            &   Mean of Core-up-low  &                &   0.56$\pm$1.39                  &   {0.61$\pm$1.51}        \\                                     
                 &            &   Core-100$\arcsec$                    &    0.55        &   1.26$\pm$2.16  &   {1.37$\pm$2.35}        \\                                     
                 &            &   Bright\_rim                          &    0.83        &   1.43$\pm$2.73  &                        \\                              
\hline
  8470 -- 8710      &    lines 1,2,3 & Core-up                 &    0.76        &   1.93$\pm$1.38  &                        \\                              
                 &            &   Core-low                &    0.76        &   2.84$\pm$1.38  &                        \\                              
                 &            &   Mean of Core-up-low  &                &   2.35$\pm$0.97                  &   {2.55$\pm$1.05}        \\                                     
                 &            &   Core-100$\arcsec$                    &    0.59        &   2.37$\pm$1.06  &   {2.58$\pm$1.15}        \\                                     
                 &            &   Bright\_rim                          &    0.92        &   1.00$\pm$1.67  &                        \\                              
   \hline
\end{tabular}
\end{minipage}
\tablefoottext{a}{For the attenuation factor the value  $h(\tau)= 0.92$ was adopted.\newline}

\begin{minipage}{180mm}

   \caption{As Table \ref{Table:results_GAIA} but using the FORS Bright\_rim spectrum as template in the fit }
\label {Table:results_Bright}
\begin{tabular}{llllcl}
\hline
\hline
Wavelength & Line(s)&FORS spectrum & rms & $h(\tau)I_{EBL} \pm \sigma_{\rm stat}$ & $I_{EBL} \pm \sigma_{\rm stat}$  \tablefootmark{a} \\
      &          &           & [cgs]&  [cgs]      &      [cgs]        \\

\hline
8470 -- 8530       &     line 1 &    Core-up                  &  0.76    &   -0.86$\pm$3.54  &                        \\                                 
                &            &    Core-low                  &  0.69    &   -6.35$\pm$3.22  &                        \\                               
                &            &    Core-$100\arcsec$                      &  0.59    &   -2.91$\pm$2.74  &                        \\                               
\hline
  8515 -- 8590  &     line 2 &    Core-up                  &  0.55    &    0.03$\pm$1.46  &                        \\                               
                &            &    Core-low                  &  0.41    &    1.56$\pm$1.10  &                        \\                               
                &            &    Mean of Core-up-low    &          &    1.02$\pm$0.88  &   1.76$\pm$1.52         \\                                 

                &            &     Core-100$\arcsec$                     &  0.35    &    0.99$\pm$0.91  &   1.71$\pm$1.57         \\                                 
\hline
  8630 -- 8690     &     line 3 &    Core-up                 &  0.52    &   -2.24$\pm$1.57  &                        \\                               
                &            &    Core-low                  &  0.90    &    3.38$\pm$2.73  &                        \\                               
                &            &    Mean of Core-up-low    &          &   -0.82$\pm$1.37  &  -1.41$\pm$2.36        \\                              -  
                &            &    Core-100$\arcsec$                      &  0.55    &    0.79$\pm$1.68  &   1.36$\pm$2.90        \\                                 
\hline
 8470 -- 8710      &   lines 1,2,3 & Core-up                  &  0.69    &    0.47$\pm$1.15  &                        \\                               
                &            &    Core-low                  &  0.77    &    1.36$\pm$1.28  &                        \\                               
                &            &    Mean of Core-up-low    &          &    0.87$\pm$0.86  &   1.50$\pm$1.48        \\                                 
                &            &    Core-100$\arcsec$                      &  0.55    &    0.87$\pm$0.92  &   1.50$\pm$1.59        \\                                 
\hline                          
\end{tabular}
\end{minipage}
\tablefoottext{a}{For the attenuation factor the value  $h(\tau)= 0.58$ was adopted.}

\end{table*}

\begin{figure*}
\sidecaption
  \includegraphics[width=12cm]{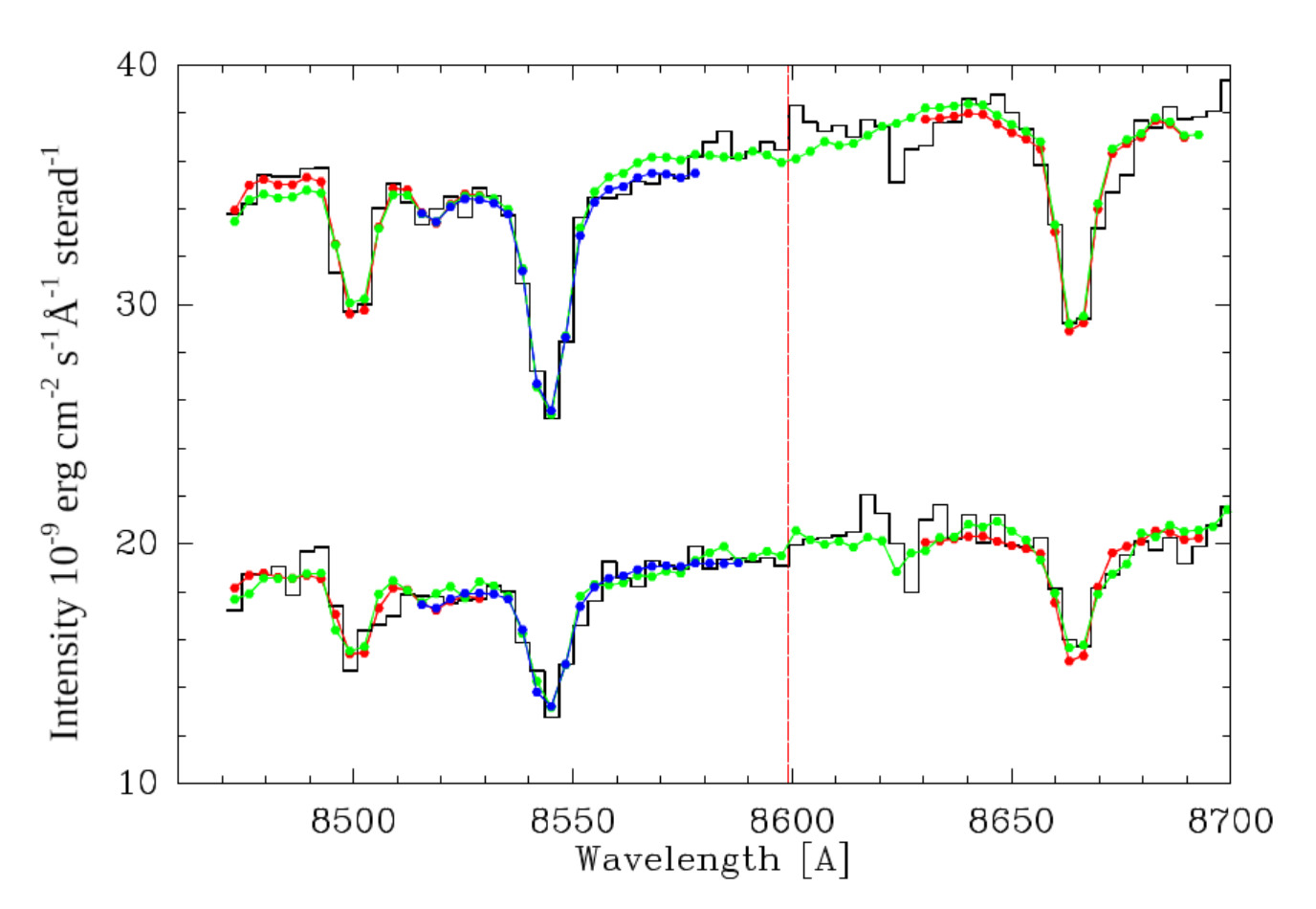}
  \caption{ Bright\_rim (upper) and Core-up (bottom) spectra, fitted
    with the Gaia integrated starlight spectrum. Individual
    \caii\ line fits for lines 1, 2, and 3 are shown in red, blue, and
    red, respectively; the joint fit of the wavelength range 8470 --
    8700\thinspace \AA\,, covering all three lines, is in green.  \newline \newline}
  \label{Fig_4}
 \end{figure*}

\begin{figure}
\vspace{-10mm}
\hspace{-8mm}
\vspace{-10mm}
      \includegraphics[width=85mm, angle=-90]{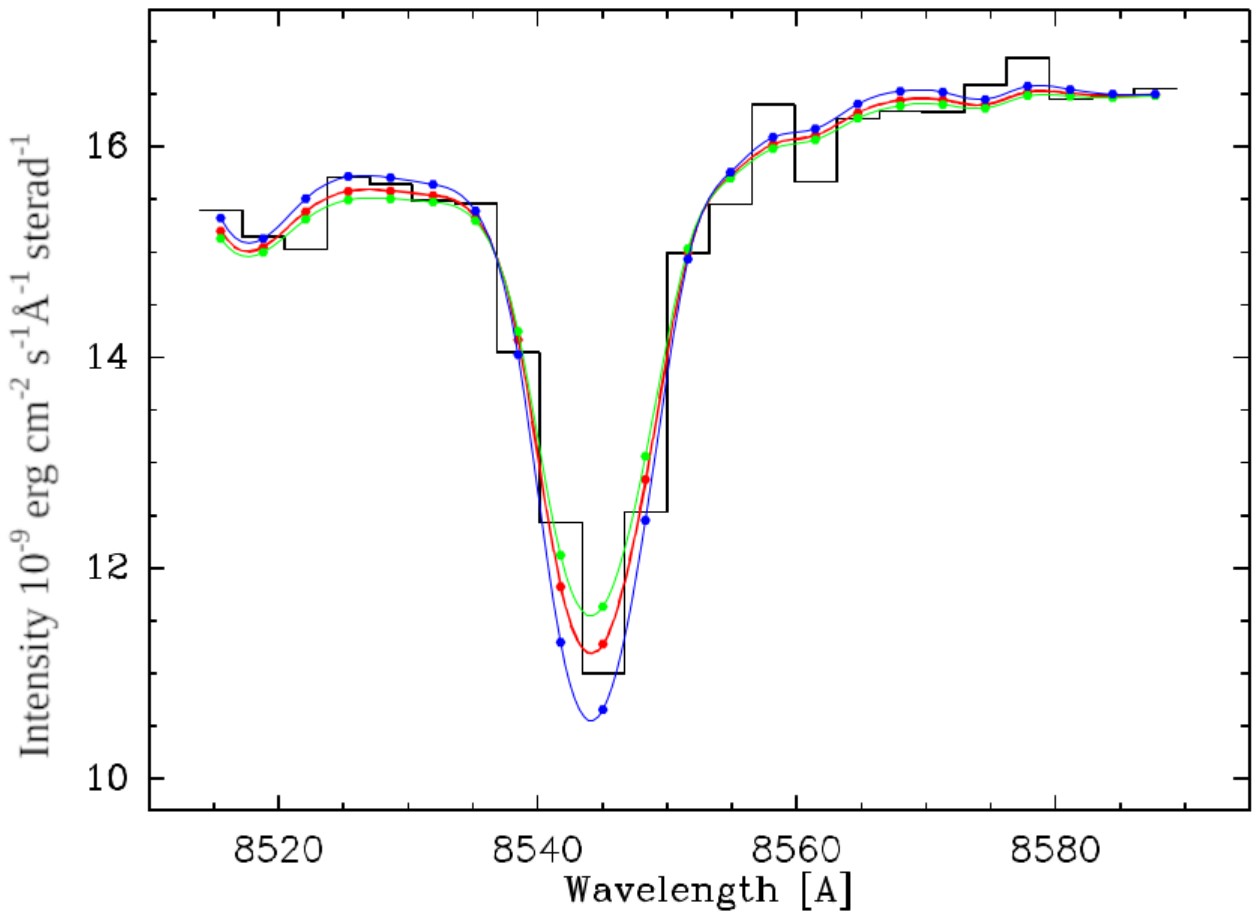}
      \caption{ Line 2 (8542\thinspace \AA\,) in the
          Core\_100\arcsec\, spectrum fitted with the Gaia integrated
        starlight spectrum. Fits are shown for assumed EBL intensities
        of $h(\tau)I_{\rm EBL}$\, = 0, 1.5, and 4.1 \cgs\,, in green,
        red, and blue, respectively.  A cubic spline interpolation was
        used for plotting the fitted curves.}
   \label{dark_core_line2}
 \end{figure}

\section{Results} \label{sect:results}

In order to extract from the observed dark-core spectra measures of
the EBL intensity, $I_{EBL}$, we have fitted the \caiitr\ lines in
accordance with Eq. \ref{Eq7}.  We have used two choices for the
fitting template: first, the GAIA RVS integrated-starlight spectrum;
second, our observed FORS spectrum of the bright rim.  We present the results of the fits in Tables \ref
{Table:results_GAIA} and \ref {Table:results_Bright}, separately for
each of the lines, 1, 2, and 3 and also for an overall fit covering
the whole wavelength range 8470 - 8710\thinspace \AA \,.

We consider first line 2 (8542\thinspace \AA\,) for which the most
conclusive results can be achieved.  Results for three different
spectral samples, Core-up, Core-low, and Core-100\arcsec are given
(see Fig. \ref{figure:slits} and Table \ref{Table:slitpos} for their
positioning along the slit).  The spectra Core-up and Core-low provide
measures that are independent of each other; the spectrum
Core-100\arcsec, encompassing them both is seen to be, as expected,
closely equal to their mean.  This is seen to be true for both fitting
methods, that is using either the GAIA RVS starlight sum or the
observed { Bright\_rim spectrum} as the template.  The two
approaches are complementary, each having its advantages and
disadvantages.  The statistical errors of the GAIA RVS spectrum are
clearly much smaller than those for the
{ Bright\_rim spectrum.  The
Bright\_rim spectrum} is noisier, however, because of its obvious
observational errors, a better representative of the total incident
radiation field spectrum. Both the dark core and the bright rim are
exposed to the same incident radiation field. Because of the different
opacities their scattered light continuum spectra may have different
slopes, but the depths and shapes of the \caiitr\ lines will remain
the same.  Combining the results for { Core-up, Core-low, and
  Core-100\arcsec,} our best estimates for $I_{EBL}$ are
$I_{EBL}=1.62\pm0.76$\thinspace cgs when using the GAIA template, and
$I_{EBL}= 1.74\pm1.52$\thinspace cgs} when using the FORS bright-rim
template. They are seen to be in complete agreement with each other, 
but because they are not independent of each other it is not
meaningful to average them. However, their agreement gives support for
the validity of each of the two approaches.  We adopt the GAIA-fitting
value, { $I_{EBL}=1.62\pm0.76$\thinspace cgs, as our best estimate}.

The weakness of line 1 (8498\thinspace \AA\,) relative to the noise
prevents a useful result from being achieved.  In the case of line 3
(8662\thinspace \AA \,), the wavelength domain is strongly disturbed,
even after the correction described in Section \ref{reduction}, by
residuals of the broad rapidly variable airglow feature of O$2$(0-1)
(see Fig. \ref{figure:noise}). As a consequence, the statistical
errors of the $I_{EBL}$ estimates are two to three times as large as
those for line 2. The best estimates obtained using the GAIA RVS and
the { Bright\_rim } fitting templates are { are $0.82\pm1.51$ and
  $-0.30\pm2.35$\,cgs,} respectively. Within their large error bars
the estimates do not disagree with the results from line 2.  Because
of the substantially smaller errors and the good agreement between the
GAIA RVS and the { Bright\_rim } based values, we rank the line 2
results as superior to those obtained from line 3.

\begin{figure}
\vspace{-10mm}
\hspace{-6mm}
\vspace{-5mm}
      \includegraphics[width=100mm]{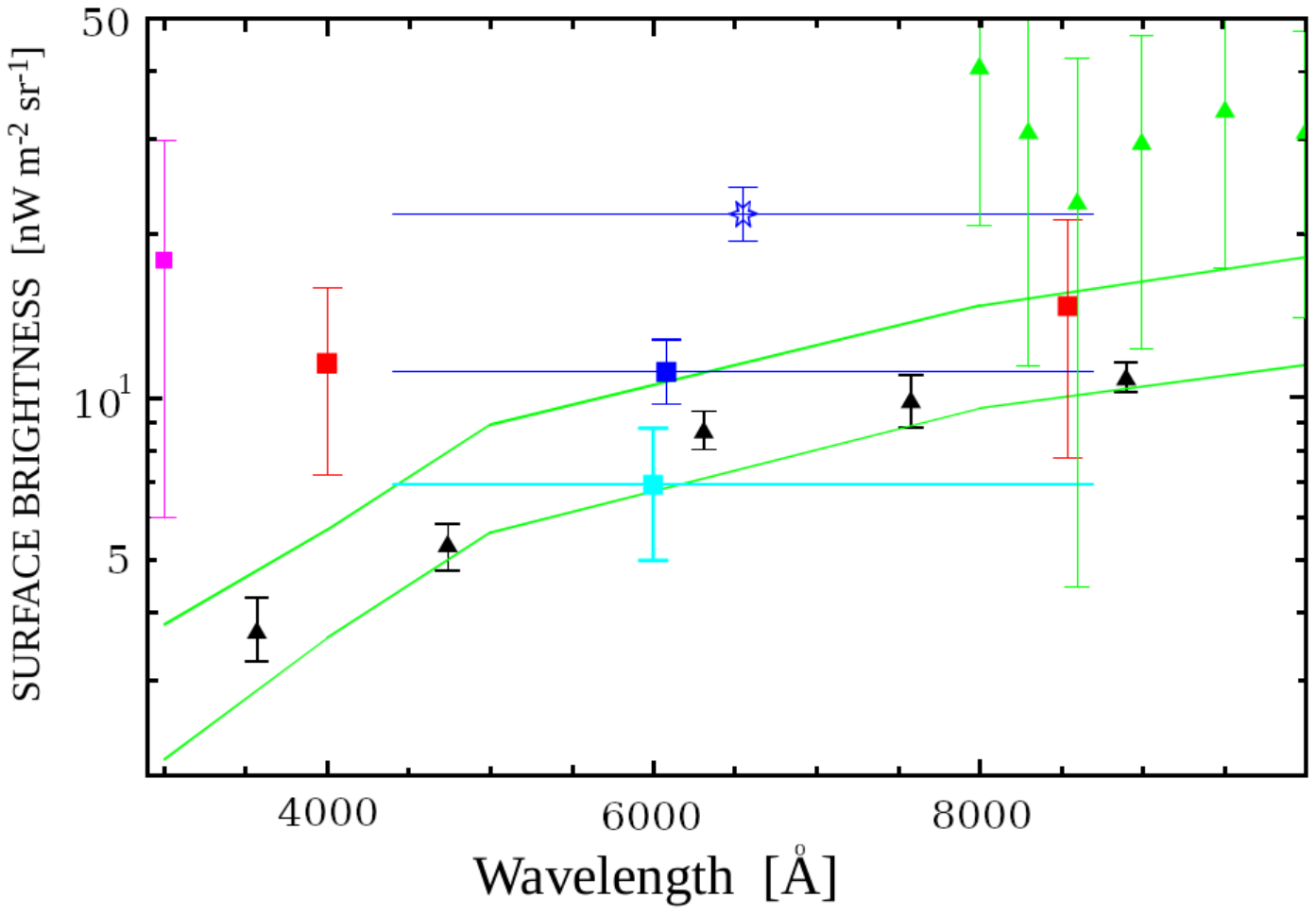}
\caption{Some recent results for EBL between 3000\thinspace \AA\, and
  1.2 $\mu$m.  Black triangles: Integrated galaxy light
  \citep{tompkins2025}.  Red squares with $1\sigma$ error bars:
  Photometric EBL measurements (at 8542\thinspace \AA, this paper) and
  \citep[at 4000\thinspace \AA,][]{mattila_vaisanen_etal_2017}, green
  triangles: \citet{matsuuraetal2017}, magenta square:
  \citet{bernstein07}, Dark blue square: New Horizons/LORRI
  measurements (the range of the dark blue bar 4400 - 8700 \AA\,
  indicates the bandpass of LORRI), blue asterisk:
  \citet{symonsetal2023}, blue square: \citet{postman24}, green lines:
  EBL from $\gamma$-ray attenuation, upper and lower limits from
  Fermi-LAT Large Area Telescope \citep{fermi18}, light blue square
  (the wavelength range was chosen to be the same as the New
  Horizons/LORRI pass band): \citet{biteau24}}
\label{comb}
\end{figure}

As our final result for the EBL intensity we adopt the value obtained
for line 2: { $I_{EBL}(\lambda)=1.62\pm0.76 (\sigma_{\rm
    stat})$\,\cgs\,. This corresponds to
  $\lambda\,I_{EBL}(\lambda)=13.83\pm6.49 (\sigma_{\rm
    stat})$\,\nW\,.}  In Fig. \ref{comb} we show this result together
with a number of other $\lambda\,I_{EBL}(\lambda)$ values in the
visible and NIR from recent literature.

\section{Discussion} \label{sect:discussion}

A direct approach to EBL measurement is to observe the total dark
night sky brightness and to model or make dedicated observations of
all foreground components. For a ground-based observer they include
the airglow, the zodiacal light, the (unresolved) starlight and the
diffuse galactic light (DGL, light scattered by dust).  The problem is
further complicated because light has to pass through the atmosphere
before reaching the telescope: each brightness component not only
suffers extinction, but is also scattered by the molecular and aerosol
components, each with its own scattering function (see
\citealt{ashburn54, mattila03, bernstein05}).

\subsection{EBL measurement with the dark cloud shadow method: Tentative detection at 8542 \AA\,}
Our observing method in the present paper is a differential one: the
telescope is switched in rapid succession between an {\em on} position
in the dark core of \DCL\, and an {\em off} { position} outside the
cloud.  The method is based on measurement of the {\em on--off}\,
surface brightness difference spectrum between nearby positions on the
sky. Our {\em on} and {\em off} positions are at slightly {different
  ecliptic latitudes}. Because the \caiitr\ lines are strong in the
solar spectrum, this could cause a small systematic offset in their
strengths in the {\em on--off}\, spectrum. Similarly, they are at
slightly {different zenith distances}, and thus the airglow's zenith
distance dependence could have an effect. However, the spatial
separation between {\em on} and {\em off}\ is only 300\arcsec\, (see
Section \ref{strategy}), small enough so that the effect caused by
these spatial gradients is of no importance.  In this respect the dark
cloud shadow method is equivalent to a night sky measurement carried
out from a location outside the atmosphere, the zodiacal cloud, and
even beyond any uniform surface-brightness layer, such as an
interstellar dust sheet between us and the globule. The foreground
components, 10--100 times brighter than the EBL, cause a substantial
photon noise.  In addition, as we discuss in Section \ref{sect:Obs}, a
substantial statistical uncertainty may be caused by the rapid time
variability of the airglow.

The main challenge for the dark cloud method is the separation of the
light scattered by dust particles in the cloud. To suppress its level
the cloud core should have a high opacity. Although \DCL\, has a very
substantial opacity, $\tau(8500)\ga 10$, the scattered light still
dominates over the EBL intensity by a factor of 10 or more.

In addition to the scattered light there is another diffuse surface
brightness component ascribed to interstellar dust particles, the
extended red emission (ERE) \citep{witt20}. It has a continuous
spectrum, covering parts or the whole wavelength range of $\sim 5000 -
8000$\thinspace\AA\,.  Its presence in the general diffuse medium was
detected by \citet{Gordon98} and has been recently confirmed by
\citet{chellew2022}. An indication of its presence in \DCL\,, and the
nearby TPN globule was also detected \citet{lehtinen_mattila_2013};
however, its presence in \DCL\, was limited to
$\lambda<8000$\thinspace\AA\,.  The \caiitr\ lines at 8500 -
8660\thinspace\AA\, are thus beyond the ERE wavelength range.
     
The \caiitr\ lines offer a very good indicator for and a robust
measure of the scattered starlight and its intensity. The depths of
the lines in the globule's spectrum can be well fitted with a model
spectrum of the incident starlight. The noise of the observed {\em
  on--off}\, FORS spectra determines the level of the statistical
error of the resulting $I_{EBL}$. The noise includes both the photon
statistics and the effect of the AGL short-term time variation. The
transfer of a FORS spectrum's noise to the statistical error of
$I_{EBL}$ is governed by the least-squares fitting procedure
MPFITFUN. The rms noise for each spectrum and the resulting
$\sigma_{\rm stat}$ value for $I_{\rm EBL}$ are given in Tables \ref
{Table:results_GAIA} and \ref{Table:results_Bright} for the individual
spectral slots surrounding lines 1, 2, and 3, and for the whole
spectral range 8470 -- 8710\thinspace \AA\,.

A systematic uncertainty of $I_{EBL}$ is caused because the
calibration of an extended surface brightness is different from that
for a point source, such as a star.  This is due to the aperture
correction, $T(A)$. $T(A)=1$ for a uniform extended surface
brightness, such as the EBL or the widely distributed scattered light
in the globule area. For observations of stars, $T(A)$ can be
substantially smaller than 1, however; its uncertainty for our
standard star observations with FORS2 introduces a $\pm8\%$ systematic
uncertainty to $I_{EBL}$ (see Appendix \ref{App:Cal_aperture}).
Another systematic error is caused by the uncertainty of the
attenuation factor, $h_{\tau}$ or $h_{\rm eff}$. It has been estimated
to be $\pm2\%$ for each of the two cases: with either GAIA\_RVS or the
observed Bright\_rim spectrum used as template to fit the Dark\_core
spectrum (see Appendix \ref{App:shadowing}).  The total absolute
uncertainty (scaling factor) is thus estimated to be $\pm10\%$.

\subsection{Other measurements of the EBL: Recent results and outlook}
 The EBL results from the {CIBER} rocket experiment
 \citep{matsuuraetal2017} cover the wavelength range of our present
 study well. Their nominal EBL value at 8600 \AA\, is given as
 $\lambda\,I_{EBL}(\lambda)=23.1\pm3.2(\sigma_{stat})$ +15.1/-14.4
 $(\sigma_{syst})$ \nW\, (their Table 3).  It is, within their and our
 error estimates, consistent with our EBL value at 8542\thinspace \AA\,.
 Recently, dark night sky measurements have been made with the LORRI
 experiment aboard NASA's New Horizons mission from locations in the
 outer Solar System, out to 57 AU.  Two teams \citep{symonsetal2023,
   postman24} have made use of the instrument's capabilities to
 measure the EBL intensity. Their most recent results are displayed in
 Fig. \ref {comb}.

Our $I_{EBL}$ value seemingly falls halfway between these two
LORRI-based values.  The comparison is problematic, however, since the
LORRI band pass covers the very broad wavelength range of 4400 --
8700\thinspace\AA\, and our narrow-band measurement falls at the very
red end of LORRI's pass band where its response is
low. Thus, even a substantially smaller or larger narrow band value of
$I_{EBL}$ at 8600\thinspace \AA\, would have only a minor effect on LORRI's
broad-band $I_{EBL}$ value. Our $I_{EBL}$ value is compatible, within
its error limits, with the integrated galaxy light (IGL) estimate of
\citet{tompkins2025} without any contribution by an additional diffuse
EBL component.  On the other hand, our result is, within its error
bars, consistent with a substantially larger $I_{EBL}$ value, in
agreement with the \citep{matsuuraetal2017} results. These two
independent results might indicate an upturn of the EBL intensity in
the red/near-IR domain at $\lambda \ga 7500$\thinspace\AA\,.  Such an
upturn, if steep enough, would remain undetectable in the broad-band
LORRI result.

 Recently, \citet{OBrien2025} using extensive 0.4--1.6 $\mu$m HST
 SKYSURF data, presented a new zodiacal light model. Based on this
 model, they announced the detection of an excess of diffuse light of
 $0.013\pm0.006$ MJy\,sterad$^{-1}$ or $\sim$45 \nW, which they
 suggested most likely originates from a spherical inner Solar System
 zodiacal dust cloud.  If instead it were of extragalactic origin, it
 would strongly exceed the upper bounds set by the broad-band LORRI
 mesurements and also our narrow-band EBL value at
 8542\thinspace\AA\,.

The diffuse Galactic light was the most important astrophysical
foreground component in LORRI's fields that had to be analysed and
subtracted. In that respect the situation is similar to the dark cloud
method where the scattered light from dust represented the main task
of our analysis.  In principle, the dark cloud shadow method has the
advantage that it is capable of narrow-band and even of spectroscopic
measurements. However, the integrated starlight spectrum does not
possess many other spectral features as good as
the \caiitr\ lines that could be utilised for the scattered-light
subtraction.

At present, improved observational techniques could already be
realised via the dark cloud shadow method. These include the
following: (1) a search for better (i.e. darker) target globules using
the very deep low surface brightness surveys becoming available (the
best targets would be those in the outskirts or even outside our
Galaxy; see \citealt{Park1998}); (2) larger telescopes to enable a
better spectral resolution, crucial for utilising deep absorption
lines such as the \caiitr\ lines, which offer a very good indicator
for and a robust measure of the scattered starlight and its
intensity. The depths of the lines in the globule's spectrum can be
well fitted with a model spectrum of the incident starlight. The noise
of the observed {\em on--off}\, FORS spectra determines the level of
the statistical error of the resulting $I_{EBL}$. The noise includes
both the photon statistics as well as the effect of the AGL short-term
time variation. The transfer of a FORS spectrum's noise to the
statistical error of $I_{EBL}$ is governed by the least-squares
fitting procedure MPFITFUN. The rms noise for each spectrum and the
resulting $\sigma_{\rm stat}$ value for $I_{\rm EBL}$ are given in
Tables \ref {Table:results_GAIA}  and \ref {Table:results_Bright}for the individual spectral slots
surrounding lines 1, 2, and 3, and for the whole spectral range 8470
-- 8710 \AA\,.

A systematic uncertainty of $I_{EBL}$ is caused because the
calibration of an extended surface brightness is different from that
for a point source, such as a star.  This is due to the aperture
correction, $T(A)$. $T(A)=1$ for a uniform extended surface
brightness, such as the EBL or the widely distributed scattered light
in the globule area. For observations of stars, $T(A)$ can be
substantially smaller than 1; however, its uncertainty for our
standard star observations with FORS2 introduces a $\pm8\%$ systematic
uncertainty to $I_{EBL}$ (see Appendix \ref{App:Cal_aperture}).
Another systematic error is caused by the uncertainty of the
attenuation factor, $h_{\tau}$ or $h_{\rm eff}$. It has been estimated
to be $\pm2\%$ for each of the two cases: with either GAIA\_RVS or the
observed Bright\_rim spectrum used as template to fit the Core-up
spectrum (see Appendix \ref{App:Cal_aperture}).  The total absolute
uncertainty (scaling factor) is thus estimated to be $\pm10\%$.

\section{Conclusions} \label{sect:concl}  

Our EBL measurement at 8542\thinspace\AA\,,
$\lambda\,I_{EBL}(\lambda)={13.8\pm6.5} (\sigma_{\rm stat})$ \nW\,
represents a tentative detection at the $2\sigma$ level. The
systematic error is estimated to be $\pm10\%$.  This value, while
compatible with the integrated galaxy light (IGL) in the z band
\citep{tompkins2025}, still allows a contribution by unresolved
sources or diffuse light of unknown origin to the EBL.  Because it is
essentially monochromatic, our $I_{\rm EBL}$ value is complementary to
the recent New Horizons/LORRI \citep{symonsetal2023, postman24} and
blazar gamma-ray attenuation \citep{biteau24}\, $I_{EBL}$ values,
which were based on broad-band measurements in the range $\sim4400 -
8700$\, \AA\,.

\begin{acknowledgements} 

This research has made use of the following resources: data products
from observations made with ESO Telescopes at the La Silla or Paranal
Observatories under ESO programmes ID 099.A-0028, 0102.A-0280 and
0104.A-0192; FORS pipeline for data reduction and quality control of the
FORS instrument data, as described in VLT-MAN-ESO-19500-4106;
NASA’s Astrophysics Data System Bibliographic Services;
the European Space Agency (ESA) space mission Gaia. Gaia data are
being processed by the Gaia Data Processing and Analysis Consortium
(DPAC). Funding for the DPAC is provided by national institutions, in
particular the institutions participating in the Gaia MultiLateral
Agreement (MLA). The Gaia mission website is
https://www.cosmos.esa.int/gaia. The Gaia archive website is
https://archives.esac.esa.int/gaia; IRAF is distributed
by the National Optical Astronomy Observatories, which are operated by
the Association of Universities for Research in Astronomy, Inc., under
cooperative agreement with the National Science Foundation.
\end{acknowledgements}

\bibliographystyle{aa} 
\bibliography{aa57566-25.bib}

@ARTICLE{appenzeller98,
       author = {{Appenzeller}, I. and {Fricke}, K. and {F{\"u}rtig}, W. and {G{\"a}ssler}, W. and {H{\"a}fner}, R. and {Harke}, R. and {Hess}, H. -J. and {Hummel}, W. and {J{\"u}rgens}, P. and {Kudritzki}, R. -P. and {Mantel}, K. -H. and {Meisl}, W. and {Muschielok}, B. and {Nicklas}, H. and {Rupprecht}, G. and {Seifert}, W. and {Stahl}, O. and {Szeifert}, T. and {Tarantik}, K.},
        title = "{Successful commissioning of FORS1 - the first optical instrument on the VLT.}",
      journal = {The Messenger},
         year = 1998,
        month = dec,
       volume = {94},
        pages = {1-6},
       adsurl = {https://ui.adsabs.harvard.edu/abs/1998Msngr..94....1A}
}

@ARTICLE{ashburn54,
       author = {{Ashburn}, Edward V.},
        title = "{The Effect of Atmospheric Scattering and Ground Reflection upon the Determination of the Height of the Night Airglow}",
      journal = {\jgr},
         year = 1954,
        month = mar,
       volume = {59},
       number = {1},
        pages = {67-70},
          doi = {10.1029/JZ059i001p00067},
       adsurl = {https://ui.adsabs.harvard.edu/abs/1954JGR....59...67A}
}

@ARTICLE{OBrien2025,
       author = {{O'Brien}, Rosalia and {Arendt}, Richard G. and {Windhorst}, Rogier A. and {Acharya}, Tejovrash and {Calamida}, Annalisa and {Carleton}, Timothy and {Carter}, Delondrae and {Cohen}, Seth H. and {Dwek}, Eli and {Frye}, Brenda L. and {Jansen}, Rolf A. and {Kenyon}, Scott J. and {Koekemoer}, Anton M. and {MacKenty}, John and {Miller}, Megan and {Ortiz}, III, Rafael and {Smith}, Peter C.~B. and {Tompkins}, Scott A.},
        title = "{SKYSURF-11: A New Zodiacal Light Model Optimized for Optical Wavelengths}",
      journal = {arXiv e-prints},
         year = 2025,
        month = oct,
          eid = {arXiv:2510.18231},
        pages = {arXiv:2510.18231},
          doi = {10.48550/arXiv.2510.18231},
archivePrefix = {arXiv},
       eprint = {2510.18231},
 primaryClass = {astro-ph.CO},
       adsurl = {https://ui.adsabs.harvard.edu/abs/2025arXiv251018231O}
}

@ARTICLE{tompkins2025,
       author = {{Tompkins}, Scott A. and {Driver}, Simon P. and {Robotham}, Aaron S.~G. and {Windhorst}, Rogier A. and {Carter}, Delondrae and {Carleton}, Timothy and {Goisman}, Zak and {Henningsen}, Daniel and {Davies}, Luke J. and {Bellstedt}, Sabine and {D'Silva}, Jordan C.~J. and {Li}, Juno and {Cohen}, Seth H. and {Jansen}, Rolf A. and {O'Brien}, Rosalia and {Koekemoer}, Anton M. and {Grogin}, Norman and {MacKenty}, John},
        title = "{SKYSURF IX -- The Cosmic Optical and Infrared Background from Integrated Galaxy Light Measurements}",
      journal = {arXiv e-prints},
         year = 2025,
        month = jul,
          eid = {arXiv:2507.03412},
        pages = {arXiv:2507.03412},
          doi = {10.48550/arXiv.2507.03412},
archivePrefix = {arXiv},
       eprint = {2507.03412},
 primaryClass = {astro-ph.CO},
       adsurl = {https://ui.adsabs.harvard.edu/abs/2025arXiv250703412T}
}

@ARTICLE{witt1974,
       author = {{Witt}, A.~N. and {Stephens}, T.~C.},
        title = "{Monte Carlo calculations of the surface brightness profiles of spherical dark nebulae}",
      journal = {\aj},
         year = 1974,
        month = sep,
       volume = {79},
        pages = {948},
          doi = {10.1086/111635},
       adsurl = {https://ui.adsabs.harvard.edu/abs/1974AJ.....79..948W}
}

@ARTICLE{bernstein05,
       author = {{Bernstein}, Rebecca A. and {Freedman}, Wendy L. and {Madore}, Barry F.},
        title = "{Corrections of Errors in ``The First Detections of the Extragalactic Background Light at 3000, 5500, and 8000 {\r{A}}. I, II, and III'' (ApJ, 571; 56, 85, 107 [2002])}",
      journal = {\apj},
         year = 2005,
        month = oct,
       volume = {632},
       number = {2},
        pages = {713-717},
          doi = {10.1086/444488},
archivePrefix = {arXiv},
       eprint = {astro-ph/0507033},
 primaryClass = {astro-ph},
       adsurl = {https://ui.adsabs.harvard.edu/abs/2005ApJ...632..713B}
}

@ARTICLE{bernstein07,
       author = {{Bernstein}, R.~A.},
        title = "{The Optical Extragalactic Background Light. Revisions and Further Comments}",
      journal = {\apj},
         year = 2007,
       volume = {666},
        pages = {663-673}
}

@ARTICLE{biteau24,
       author = {{Gr{\'e}aux}, Lucas and {Biteau}, Jonathan and {Nievas Rosillo}, Mireia},
        title = "{The Cosmological Optical Convergence: Extragalactic Background Light from TeV Gamma Rays}",
      journal = {\apjl},
         year = 2024,
        month = nov,
       volume = {975},
       number = {1},
          eid = {L18},
        pages = {L18},
          doi = {10.3847/2041-8213/ad85c9},
archivePrefix = {arXiv},
       eprint = {2410.07011},
 primaryClass = {astro-ph.HE},
       adsurl = {https://ui.adsabs.harvard.edu/abs/2024ApJ...975L..18G}
}

@ARTICLE{bruzual03,
       author = {{Bruzual}, G. and {Charlot}, S.},
        title = "{Stellar population synthesis at the resolution of 2003}",
      journal = {\mnras},
         year = 2003,
        month = oct,
       volume = {344},
       number = {4},
        pages = {1000-1028},
          doi = {10.1046/j.1365-8711.2003.06897.x},
archivePrefix = {arXiv},
       eprint = {astro-ph/0309134},
 primaryClass = {astro-ph},
       adsurl = {https://ui.adsabs.harvard.edu/abs/2003MNRAS.344.1000B}
}

@ARTICLE{carnall2017,
       author = {{Carnall}, A.~C.},
        title = "{SpectRes: A Fast Spectral Resampling Tool in Python}",
      journal = {arXiv e-prints},
         year = 2017,
        month = may,
          eid = {arXiv:1705.05165},
        pages = {arXiv:1705.05165},
          doi = {10.48550/arXiv.1705.05165},
archivePrefix = {arXiv},
       eprint = {1705.05165},
 primaryClass = {astro-ph.IM},
       adsurl = {https://ui.adsabs.harvard.edu/abs/2017arXiv170505165C}
}

@ARTICLE{chellew2022,
       author = {{Chellew}, Blake and {Brandt}, Timothy D. and {Hensley}, Brandon S. and {Draine}, Bruce T. and {Matthaey}, Eve},
        title = "{An Optical Spectrum of the Diffuse Galactic Light from BOSS and IRIS}",
      journal = {\apj},
         year = 2022,
        month = jun,
       volume = {932},
       number = {2},
          eid = {112},
        pages = {112},
          doi = {10.3847/1538-4357/ac6efc},
archivePrefix = {arXiv},
       eprint = {2201.01378},
 primaryClass = {astro-ph.GA},
       adsurl = {https://ui.adsabs.harvard.edu/abs/2022ApJ...932..112C}
}

@ARTICLE{driveretal2016,
       author = {{Driver}, Simon P. and {Andrews}, Stephen K. and {Davies}, Luke J. and {Robotham}, Aaron S.~G. and {Wright}, Angus H. and {Windhorst}, Rogier A. and {Cohen}, Seth and {Emig}, Kim and {Jansen}, Rolf A. and {Dunne}, Loretta},
        title = "{Measurements of Extragalactic Background Light from the Far UV to the Far IR from Deep Ground- and Space-based Galaxy Counts}",
      journal = {\apj},
         year = 2016,
        month = aug,
       volume = {827},
       number = {2},
          eid = {108},
        pages = {108},
          doi = {10.3847/0004-637X/827/2/108},
archivePrefix = {arXiv},
       eprint = {1605.01523},
 primaryClass = {astro-ph.GA},
       adsurl = {https://ui.adsabs.harvard.edu/abs/2016ApJ...827..108D}
}

@ARTICLE{flynn06,
       author = {{Flynn}, Chris and {Holmberg}, Johan and {Portinari}, Laura and {Fuchs}, Burkhard and {Jahrei{\ss}}, Hartmut},
        title = "{On the mass-to-light ratio of the local Galactic disc and the optical luminosity of the Galaxy}",
      journal = {\mnras},
         year = 2006,
        month = nov,
       volume = {372},
       number = {3},
        pages = {1149-1160},
          doi = {10.1111/j.1365-2966.2006.10911.x},
archivePrefix = {arXiv},
       eprint = {astro-ph/0608193},
 primaryClass = {astro-ph},
       adsurl = {https://ui.adsabs.harvard.edu/abs/2006MNRAS.372.1149F}
}

@ARTICLE{Gordon98,
       author = {{Gordon}, Karl D. and {Witt}, Adolf N. and {Friedmann}, Brian C.},
        title = "{Detection of Extended Red Emission in the Diffuse Interstellar Medium}",
      journal = {\apj},
         year = 1998,
        month = may,
       volume = {498},
       number = {2},
        pages = {522-540},
          doi = {10.1086/305571},
archivePrefix = {arXiv},
       eprint = {astro-ph/9712032},
 primaryClass = {astro-ph},
       adsurl = {https://ui.adsabs.harvard.edu/abs/1998ApJ...498..522G}
}

@INPROCEEDINGS{gordon_2004,
       author = {{Gordon}, K.~D.},
        title = "{Interstellar Dust Scattering Properties}",
    booktitle = {Astrophysics of Dust},
         year = 2004,
       editor = {{Witt}, Adolf N. and {Clayton}, Geoffrey C. and {Draine}, Bruce T.},
       series = {Astronomical Society of the Pacific Conference Series},
       volume = {309},
        month = may,
        pages = {77},
          doi = {10.48550/arXiv.astro-ph/0309709},
archivePrefix = {arXiv},
       eprint = {astro-ph/0309709},
 primaryClass = {astro-ph},
       adsurl = {https://ui.adsabs.harvard.edu/abs/2004ASPC..309...77G}
}

@ARTICLE{hauseretal_1998,
       author = {{Hauser}, M.~G. and {Arendt}, R.~G. and {Kelsall}, T. and {Dwek}, E. and {Odegard}, N. and {Weiland}, J.~L. and {Freudenreich}, H.~T. and {Reach}, W.~T. and {Silverberg}, R.~F. and {Moseley}, S.~H. and {Pei}, Y.~C. and {Lubin}, P. and {Mather}, J.~C. and {Shafer}, R.~A. and {Smoot}, G.~F. and {Weiss}, R. and {Wilkinson}, D.~T. and {Wright}, E.~L.},
        title = "{The COBE Diffuse Infrared Background Experiment Search for the Cosmic Infrared Background. I. Limits and Detections}",
      journal = {\apj},
         year = 1998,
        month = nov,
       volume = {508},
       number = {1},
        pages = {25-43},
          doi = {10.1086/306379},
archivePrefix = {arXiv},
       eprint = {astro-ph/9806167},
 primaryClass = {astro-ph},
       adsurl = {https://ui.adsabs.harvard.edu/abs/1998ApJ...508...25H}
}

@ARTICLE{henyey,
       author = {{Henyey}, L.~G. and {Greenstein}, J.~L.},
        title = "{Diffuse radiation in the Galaxy.}",
      journal = {\apj},
         year = 1941,
        month = jan,
       volume = {93},
        pages = {70-83},
          doi = {10.1086/144246},
       adsurl = {https://ui.adsabs.harvard.edu/abs/1941ApJ....93...70H}
}

@ARTICLE{kainulainenetal2007,
       author = {{Kainulainen}, J. and {Lehtinen}, K. and {V{\"a}is{\"a}nen}, P. and {Bronfman}, L. and {Knude}, J.},
        title = "{A comparison of density structures of a star forming and a non-star-forming globule. <ASTROBJ>DCld303.8-14.2</ASTROBJ> and <ASTROBJ>Thumbprint nebula</ASTROBJ>}",
      journal = {\aap},
         year = 2007,
        month = mar,
       volume = {463},
       number = {3},
        pages = {1029-1037},
          doi = {10.1051/0004-6361:20066431},
archivePrefix = {arXiv},
       eprint = {astro-ph/0612221},
 primaryClass = {astro-ph},
       adsurl = {https://ui.adsabs.harvard.edu/abs/2007A&A...463.1029K}
}

@ARTICLE{laueretal2021,
       author = {{Lauer}, Tod R. and {Postman}, Marc and {Weaver}, Harold A. and {Spencer}, John R. and {Stern}, S. Alan and {Buie}, Marc W. and {Durda}, Daniel D. and {Lisse}, Carey M. and {Poppe}, A.~R. and {Binzel}, Richard P. and {Britt}, Daniel T. and {Buratti}, Bonnie J. and {Cheng}, Andrew F. and {Grundy}, W.~M. and {Hor{\'a}nyi}, Mihaly and {Kavelaars}, J.~J. and {Linscott}, Ivan R. and {McKinnon}, William B. and {Moore}, Jeffrey M. and {N{\'u}{\~n}ez}, J.~I. and {Olkin}, Catherine B. and {Parker}, Joel W. and {Porter}, Simon B. and {Reuter}, Dennis C. and {Robbins}, Stuart J. and {Schenk}, Paul and {Showalter}, Mark R. and {Singer}, Kelsi N. and {Verbiscer}, Anne J. and {Young}, Leslie A.},
        title = "{New Horizons Observations of the Cosmic Optical Background}",
      journal = {\apj},
         year = 2021,
        month = jan,
       volume = {906},
       number = {2},
          eid = {77},
        pages = {77},
          doi = {10.3847/1538-4357/abc881},
archivePrefix = {arXiv},
       eprint = {2011.03052},
 primaryClass = {astro-ph.GA},
       adsurl = {https://ui.adsabs.harvard.edu/abs/2021ApJ...906...77L}
}

@ARTICLE{laueretal2022,
       author = {{Lauer}, Tod R. and {Postman}, Marc and {Spencer}, John R. and {Weaver}, Harold A. and {Stern}, S. Alan and {Gladstone}, G. Randall and {Binzel}, Richard P. and {Britt}, Daniel T. and {Buie}, Marc W. and {Buratti}, Bonnie J. and {Cheng}, Andrew F. and {Grundy}, W.~M. and {Hor{\'a}nyi}, Mihaly and {Kavelaars}, J.~J. and {Linscott}, Ivan R. and {Lisse}, Carey M. and {McKinnon}, William B. and {McNutt}, Ralph L. and {Moore}, Jeffrey M. and {N{\'u}{\~n}ez}, J.~I. and {Olkin}, Catherine B. and {Parker}, Joel W. and {Porter}, Simon B. and {Reuter}, Dennis C. and {Robbins}, Stuart J. and {Schenk}, Paul M. and {Showalter}, Mark R. and {Singer}, Kelsi N. and {Verbiscer}, Anne. J. and {Young}, Leslie A.},
        title = "{Anomalous Flux in the Cosmic Optical Background Detected with New Horizons Observations}",
      journal = {\apjl},
         year = 2022,
        month = mar,
       volume = {927},
       number = {1},
          eid = {L8},
        pages = {L8},
          doi = {10.3847/2041-8213/ac573d},
archivePrefix = {arXiv},
       eprint = {2202.04273},
 primaryClass = {astro-ph.GA},
       adsurl = {https://ui.adsabs.harvard.edu/abs/2022ApJ...927L...8L}
}

@ARTICLE{LeBorgne03,
       author = {{Le Borgne}, J. -F. and {Bruzual}, G. and {Pell{\'o}}, R. and {Lan{\c{c}}on}, A. and {Rocca-Volmerange}, B. and {Sanahuja}, B. and {Schaerer}, D. and {Soubiran}, C. and {V{\'\i}lchez-G{\'o}mez}, R.},
        title = "{STELIB: A library of stellar spectra at R \raisebox{-0.5ex}\textasciitilde 2000}",
      journal = {\aap},
         year = 2003,
        month = may,
       volume = {402},
        pages = {433-442},
          doi = {10.1051/0004-6361:20030243},
archivePrefix = {arXiv},
       eprint = {astro-ph/0302334},
 primaryClass = {astro-ph},
       adsurl = {https://ui.adsabs.harvard.edu/abs/2003A&A...402..433L}
}

@ARTICLE{mattila_1970,
       author = {{Mattila}, K.},
        title = "{Interpretation of the surface brightness of dark nebulae.}",
      journal = {\aap},
         year = 1970,
        month = nov,
       volume = {9},
        pages = {53},
       adsurl = {https://ui.adsabs.harvard.edu/abs/1970A&A.....9...53M}
}

@ARTICLE{lehtinen_mattila_2013,
       author = {{Lehtinen}, K. and {Mattila}, K.},
        title = "{Spectroscopy of diffuse light in dust clouds. Scattered light and the solar neighbourhood radiation field}",
      journal = {\aap},
         year = 2013,
        month = jan,
       volume = {549},
          eid = {A91},
        pages = {A91},
          doi = {10.1051/0004-6361/201220239},
archivePrefix = {arXiv},
       eprint = {1211.7282},
 primaryClass = {astro-ph.GA},
       adsurl = {https://ui.adsabs.harvard.edu/abs/2013A&A...549A..91L}
}

@ARTICLE{matsuuraetal2017,
       author = {{Matsuura}, Shuji and {Arai}, Toshiaki and {Bock}, James J. and {Cooray}, Asantha and {Korngut}, Phillip M. and {Kim}, Min Gyu and {Lee}, Hyung Mok and {Lee}, Dae Hee and {Levenson}, Louis R. and {Matsumoto}, Toshio and {Onishi}, Yosuke and {Shirahata}, Mai and {Tsumura}, Kohji and {Wada}, Takehiko and {Zemcov}, Michael},
        title = "{New Spectral Evidence of an Unaccounted Component of the Near-infrared Extragalactic Background Light from the CIBER}",
      journal = {\apj},
         year = 2017,
        month = apr,
       volume = {839},
       number = {1},
          eid = {7},
        pages = {7},
          doi = {10.3847/1538-4357/aa6843},
archivePrefix = {arXiv},
       eprint = {1704.07166},
 primaryClass = {astro-ph.GA},
       adsurl = {https://ui.adsabs.harvard.edu/abs/2017ApJ...839....7M}
}

@ARTICLE{mattila1976,
       author = {{Mattila}, K.},
        title = "{On the measurement of the extragalactic background brightness at 4000 {\r{A}}.}",
      journal = {\aap},
         year = 1976,
        month = feb,
       volume = {47},
        pages = {77-95},
       adsurl = {https://ui.adsabs.harvard.edu/abs/1976A&A....47...77M}
}

@ARTICLE{mattila_lehtinen_etal_2017,
       author = {{Mattila}, K. and {Lehtinen}, K. and {V{\"a}is{\"a}nen}, P. and {von Appen-Schnur}, G. and {Leinert}, Ch.},
        title = "{Extragalactic background light: a measurement at 400 nm using dark cloud shadow*$^{{\textdagger}}$- I. Low surface brightness spectrophotometry in the area of Lynds 1642}",
      journal = {\mnras},
         year = {2017a},
        month = sep,
       volume = {470},
       number = {2},
        pages = {2133-2151},
          doi = {10.1093/mnras/stx1295},
archivePrefix = {arXiv},
       eprint = {1705.10681},
 primaryClass = {astro-ph.CO},
       adsurl = {https://ui.adsabs.harvard.edu/abs/2017MNRAS.470.2133M}
}

@ARTICLE{mattila03,
       author = {{Mattila}, K.},
        title = "{Has the Optical Extragalactic Background Light Been Detected?}",
      journal = {\apj},
         year = 2003,
        month = jul,
       volume = {591},
       number = {1},
        pages = {119-124},
          doi = {10.1086/375182},
archivePrefix = {arXiv},
       eprint = {astro-ph/0303196},
 primaryClass = {astro-ph},
       adsurl = {https://ui.adsabs.harvard.edu/abs/2003ApJ...591..119M}
}

@ARTICLE{mattila_vaisanen_etal_2017,
       author = {{Mattila}, K. and {V{\"a}is{\"a}nen}, P. and {Lehtinen}, K. and {von Appen-Schnur}, G. and {Leinert}, Ch.},
        title = "{Extragalactic background light: a measurement at 400 nm using dark cloud shadow - II. Spectroscopic separation of the dark cloud's light, and results$^{★}$}",
      journal = {\mnras},
         year = {2017b},
        month = sep,
       volume = {470},
       number = {2},
        pages = {2152-2169},
          doi = {10.1093/mnras/stx1296},
archivePrefix = {arXiv},
       eprint = {1705.10790},
 primaryClass = {astro-ph.GA},
       adsurl = {https://ui.adsabs.harvard.edu/abs/2017MNRAS.470.2152M}
}

@ARTICLE{mattilaetal2018,
       author = {{Mattila}, K. and {Haas}, M. and {Haikala}, L.~K. and {Jo}, Y. -S. and {Lehtinen}, K. and {Leinert}, Ch. and {V{\"a}is{\"a}nen}, P.},
        title = "{Optical and UV surface brightness of translucent dark nebulae. Dust albedo, radiation field, and fluorescence emission by H$_{2}$}",
      journal = {\aap},
         year = 2018,
        month = sep,
       volume = {617},
          eid = {A42},
        pages = {A42},
          doi = {10.1051/0004-6361/201833196},
archivePrefix = {arXiv},
       eprint = {1806.06235},
 primaryClass = {astro-ph.GA},
       adsurl = {https://ui.adsabs.harvard.edu/abs/2018A&A...617A..42M}
}

@ARTICLE{symonsetal2023,
       author = {{Symons}, Teresa and {Zemcov}, Michael and {Cooray}, Asantha and {Lisse}, Carey and {Poppe}, Andrew R.},
        title = "{A Measurement of the Cosmic Optical Background and Diffuse Galactic Light Scaling from the R < 50 au New Horizons-LORRI Data}",
      journal = {\apj},
         year = 2023,
        month = mar,
       volume = {945},
       number = {1},
          eid = {45},
        pages = {45},
          doi = {10.3847/1538-4357/acaa37},
archivePrefix = {arXiv},
       eprint = {2212.07449},
 primaryClass = {astro-ph.CO},
       adsurl = {https://ui.adsabs.harvard.edu/abs/2023ApJ...945...45S}
}

@ARTICLE{Wainscoat92,
       author = {{Wainscoat}, Richard J. and {Cohen}, Martin and {Volk}, Kevin and {Walker}, Helen J. and {Schwartz}, Deborah E.},
        title = "{A Model of the 8--25 Micron Point Source Infrared Sky}",
      journal = {\apjs},
         year = 1992,
        month = nov,
       volume = {83},
        pages = {111},
          doi = {10.1086/191733},
       adsurl = {https://ui.adsabs.harvard.edu/abs/1992ApJS...83..111W}
}

@ARTICLE{zemcov2017,
       author = {{Zemcov}, Michael and {Immel}, P. and {Nguyen}, C., et al.},
        title = "{On the origin of near-infrared extragalactic background light anisotropy}",
      journal = {Nature Communications},
         year = 2017,
       volume = {8},
        pages = {15003}
}

@ARTICLE{postman24,
       author = {{Postman}, Marc and {Lauer}, Tod R. and {Parker}, Joel W. and {Spencer}, John R. and {Weaver}, Harold A. and {Shull}, J. Michael and {Stern}, S. Alan and {Brandt}, Pontus and {Conard}, Steven J. and {Gladstone}, G. Randall and {Lisse}, Carey M. and {Porter}, Simon B. and {Singer}, Kelsi N. and {Verbiscer}, Anne. J.},
        title = "{New Synoptic Observations of the Cosmic Optical Background with New Horizons}",
      journal = {\apj},
         year = 2024,
        month = sep,
       volume = {972},
       number = {1},
          eid = {95},
        pages = {95},
          doi = {10.3847/1538-4357/ad5ffc},
archivePrefix = {arXiv},
       eprint = {2407.06273},
 primaryClass = {astro-ph.GA},
       adsurl = {https://ui.adsabs.harvard.edu/abs/2024ApJ...972...95P}
}

@ARTICLE{togi_2017,
       author = {{Togi}, Aditya and {Witt}, Adolf N. and {John}, Demi St.},
        title = "{Dust properties of the cometary globule Barnard 207 (LDN 1489)}",
      journal = {\aap},
         year = 2017,
        month = sep,
       volume = {605},
          eid = {A99},
        pages = {A99},
          doi = {10.1051/0004-6361/201629414},
archivePrefix = {arXiv},
       eprint = {1707.07925},
 primaryClass = {astro-ph.GA},
       adsurl = {https://ui.adsabs.harvard.edu/abs/2017A&A...605A..99T}
}

@ARTICLE{Park1998,
       author = {{Park}, C. and {Kim}, J.},
        title = "{Diffuse Dark and Bright Objects in the Hubble Deep Field}",
      journal = {\apj},
         year = 1998,
        month = jul,
       volume = {501},
        pages = {23-31},
          doi = {10.1086/305796},
       adsurl = {https://ui.adsabs.harvard.edu/abs/1998ApJ...501...23P}
}

@ARTICLE{witt_1976,
       author = {{Fitzgerald}, M.~P. and {Stephens}, T.~C. and {Witt}, A.~N.},
        title = "{Surface brightness profiles of dark nebulae: the Thumbprint nebula in Chamaeleon.}",
      journal = {\apj},
         year = 1976,
        month = sep,
       volume = {208},
        pages = {709-717},
          doi = {10.1086/154654},
       adsurl = {https://ui.adsabs.harvard.edu/abs/1976ApJ...208..709F}
}

@ARTICLE{witt_1990,
       author = {{Witt}, Adolf N. and {Oliveri}, Marco V. and {Schild}, Rudolph E.},
        title = "{The Scattering Properties and Density Distribution of Dust in a Small Interstellar Cloud}",
      journal = {\aj},
         year = 1990,
        month = mar,
       volume = {99},
        pages = {888},
          doi = {10.1086/115381},
       adsurl = {https://ui.adsabs.harvard.edu/abs/1990AJ.....99..888W}
}

@ARTICLE{witt20,
       author = {{Witt}, Adolf N. and {Lai}, Thomas S. -Y.},
        title = "{Extended red emission: observational constraints for models}",
      journal = {\apss},
         year = 2020,
        month = mar,
       volume = {365},
       number = {3},
          eid = {58},
        pages = {58},
          doi = {10.1007/s10509-020-03766-w},
archivePrefix = {arXiv},
       eprint = {2003.06453},
 primaryClass = {astro-ph.GA},
       adsurl = {https://ui.adsabs.harvard.edu/abs/2020Ap&SS.365...58W}
}

@ARTICLE{koushan21,
       author = {{Koushan}, Soheil and {Driver}, Simon P. and {Bellstedt}, Sabine and {Davies}, Luke J. and {Robotham}, Aaron S.~G. and {Lagos}, Claudia del P. and {Hashemizadeh}, Abdolhosein and {Obreschkow}, Danail and {Thorne}, Jessica E. and {Bremer}, Malcolm and {Holwerda}, B.~W. and {Hopkins}, Andrew M. and {Jarvis}, Matt J. and {Siudek}, Malgorzata and {Windhorst}, Rogier A.},
        title = "{GAMA/DEVILS: constraining the cosmic star formation history from improved measurements of the 0.3-2.2 {\ensuremath{\mu}}m extragalactic background light}",
      journal = {\mnras},
         year = 2021,
        month = may,
       volume = {503},
       number = {2},
        pages = {2033-2052},
          doi = {10.1093/mnras/stab540},
archivePrefix = {arXiv},
       eprint = {2102.12323},
 primaryClass = {astro-ph.CO},
       adsurl = {https://ui.adsabs.harvard.edu/abs/2021MNRAS.503.2033K}
}

@ARTICLE{korngut22,
       author = {{Korngut}, P.~M. and {Kim}, M.~G. and {Arai}, T. and {Bangale}, P. and {Bock}, J. and {Cooray}, A. and {Cheng}, Y.~T. and {Feder}, R. and {Hristov}, V. and {Lanz}, A. and {Lee}, D.~H. and {Levenson}, L. and {Matsumoto}, T. and {Matsuura}, S. and {Nguyen}, C. and {Sano}, K. and {Tsumura}, K. and {Zemcov}, M.},
        title = "{Inferred Measurements of the Zodiacal Light Absolute Intensity through Fraunhofer Absorption Line Spectroscopy with CIBER}",
      journal = {\apj},
         year = 2022,
        month = feb,
       volume = {926},
       number = {2},
          eid = {133},
        pages = {133},
          doi = {10.3847/1538-4357/ac44ff},
archivePrefix = {arXiv},
       eprint = {2104.07104},
 primaryClass = {astro-ph.EP},
       adsurl = {https://ui.adsabs.harvard.edu/abs/2022ApJ...926..133K}
}

@ARTICLE{fermi18,
       author = {{Fermi-LAT Collaboration} and {Abdollahi}, S. and {Ackermann}, M. and {Ajello}, M. and {Atwood}, W.~B. and {Baldini}, L. and {Ballet}, J. and {Barbiellini}, G. and {Bastieri}, D. and {Becerra Gonzalez}, J. and {Bellazzini}, R. and {Bissaldi}, E. and {Blandford}, R.~D. and {Bloom}, E.~D. and {Bonino}, R. and {Bottacini}, E. and {Buson}, S. and {Bregeon}, J. and {Bruel}, P. and {Buehler}, R. and {Cameron}, R.~A. and {Caputo}, R. and {Caraveo}, P.~A. and {Cavazzuti}, E. and {Charles}, E. and {Chen}, S. and {Cheung}, C.~C. and {Chiaro}, G. and {Ciprini}, S. and {Cohen-Tanugi}, J. and {Cominsky}, L.~R. and {Conrad}, J. and {Costantin}, D. and {Cutini}, S. and {D'Ammando}, F. and {de Palma}, F. and {Desai}, A. and {Digel}, S.~W. and {Di Lalla}, N. and {Di Mauro}, M. and {Di Venere}, L. and {Dom{\'\i}nguez}, A. and {Favuzzi}, C. and {Fegan}, S.~J. and {Finke}, J. and {Franckowiak}, A. and {Fukazawa}, Y. and {Funk}, S. and {Fusco}, P. and {Gallardo Romero}, G. and {Gargano}, F. and {Gasparrini}, D. and {Giglietto}, N. and {Giordano}, F. and {Giroletti}, M. and {Green}, D. and {Grenier}, I.~A. and {Guillemot}, L. and {Guiriec}, S. and {Hartmann}, D.~H. and {Hays}, E. and {Helgason}, K. and {Horan}, D. and {J{\'o}hannesson}, G. and {Kocevski}, D. and {Kuss}, M. and {Larsson}, S. and {Latronico}, L. and {Li}, J. and {Longo}, F. and {Loparco}, F. and {Lott}, B. and {Lovellette}, M.~N. and {Lubrano}, P. and {Madejski}, G.~M. and {Magill}, J.~D. and {Maldera}, S. and {Manfreda}, A. and {Marcotulli}, L. and {Mazziotta}, M.~N. and {McEnery}, J.~E. and {Meyer}, M. and {Michelson}, P.~F. and {Mizuno}, T. and {Monzani}, M.~E. and {Morselli}, A. and {Moskalenko}, I.~V. and {Negro}, M. and {Nuss}, E. and {Ojha}, R. and {Omodei}, N. and {Orienti}, M. and {Orlando}, E. and {Ormes}, J.~F. and {Palatiello}, M. and {Paliya}, V.~S. and {Paneque}, D. and {Perkins}, J.~S. and {Persic}, M. and {Pesce-Rollins}, M. and {Petrosian}, V. and {Piron}, F. and {Porter}, T.~A. and {Primack}, J.~R. and {Principe}, G. and {Rain{\`o}}, S. and {Rando}, R. and {Razzano}, M. and {Razzaque}, S. and {Reimer}, A. and {Reimer}, O. and {Saz Parkinson}, P.~M. and {Sgr{\`o}}, C. and {Siskind}, E.~J. and {Spandre}, G. and {Spinelli}, P. and {Suson}, D.~J. and {Tajima}, H. and {Takahashi}, M. and {Thayer}, J.~B. and {Tibaldo}, L. and {Torres}, D.~F. and {Torresi}, E. and {Tosti}, G. and {Tramacere}, A. and {Troja}, E. and {Valverde}, J. and {Vianello}, G. and {Vogel}, M. and {Wood}, K. and {Zaharijas}, G.},
        title = "{A gamma-ray determination of the Universe's star formation history}",
      journal = {Science},
         year = 2018,
        month = nov,
       volume = {362},
       number = {6418},
        pages = {1031-1034},
          doi = {10.1126/science.aat8123},
archivePrefix = {arXiv},
       eprint = {1812.01031},
 primaryClass = {astro-ph.HE},
       adsurl = {https://ui.adsabs.harvard.edu/abs/2018Sci...362.1031F}
}

@ARTICLE{freyberg2004,
       author = {{Freyberg}, M.~J. and {Breitschwerdt}, D. and {Alves}, J.},
        title = "{Observations of the darkest regions in the sky: X-ray shadowing by the Bok globule Barnard 68}",
      journal = {\memsai},
         year = 2004,
        month = jan,
       volume = {75},
        pages = {509},
       adsurl = {https://ui.adsabs.harvard.edu/abs/2004MmSAI..75..509F}
}

@ARTICLE{Yeung2023,
       author = {{Yeung}, M.~C.~H. and {Freyberg}, M.~J. and {Ponti}, G. and {Dennerl}, K. and {Sasaki}, M. and {Strong}, A.},
        title = "{SRG/eROSITA X-ray shadowing study of giant molecular clouds}",
      journal = {\aap},
         year = 2023,
        month = aug,
       volume = {676},
          eid = {A3},
        pages = {A3},
          doi = {10.1051/0004-6361/202345867},
archivePrefix = {arXiv},
       eprint = {2306.05858},
 primaryClass = {astro-ph.GA},
       adsurl = {https://ui.adsabs.harvard.edu/abs/2023A&A...676A...3Y}
}

@ARTICLE{sandqvist1977,
       author = {{Sandqvist}, Aa.},
        title = "{More southern dark dust clouds.}",
      journal = {\aap},
         year = 1977,
        month = may,
       volume = {57},
        pages = {467-470},
       adsurl = {https://ui.adsabs.harvard.edu/abs/1977A&A....57..467S}
}

@ARTICLE{DC1986,
       author = {{Hartley}, M. and {Manchester}, R.~N. and {Smith}, R.~M. and {Tritton}, S.~B. and {Goss}, W.~M.},
        title = "{A catalogue of southern dark clouds.}",
      journal = {\aaps},
         year = 1986,
        month = jan,
       volume = {63},
        pages = {27-48},
       adsurl = {https://ui.adsabs.harvard.edu/abs/1986A&AS...63...27H}
}
\begin{appendix}

  \section{Observations and data reduction}
\label {Observations}
  The observations were conducted in years 2017, 2019, 2020, and 2021.
  Of the 27 spectroscopic observing blocks 21 were obseved between
  airmasses 1.65 to 1.74 and only three blocks were observed at
  airmasses 1.83 to 1.94. The minimum airmass reached for DC303
  (Decl. = $-77\degr$) at Paranal is 1.65. The number of observed
  blocks in months each year are listed in Table \ref{Table:nblocks}.

  \begin{table} [h]
\begin{minipage}{175mm}
\caption[]{Number of observing blocks observed each year and month.}
\begin{tabular}{l|c|c|c|c|c}
  \hline
  \hline
   Year &  Jan.  & Feb. & Mar. &  Jun. & Jul. \\    
  \hline 
2017  &   & 4\tablefootmark{a} &    & 4   &   \\
2019  & 1  &                    & 5  &    &   \\
2020  &    & 1                  & 9  &    &   \\
2021  & 1  &                    & 4  &    & 2 \\
\hline                          
\end{tabular}
\label {Table:nblocks}
\end{minipage}
\tablefoottext{a}{Pre-imaging}

\end{table}

\subsection {Airglow}
\label {agl}
\begin{figure} [h]
\includegraphics[width=88mm]{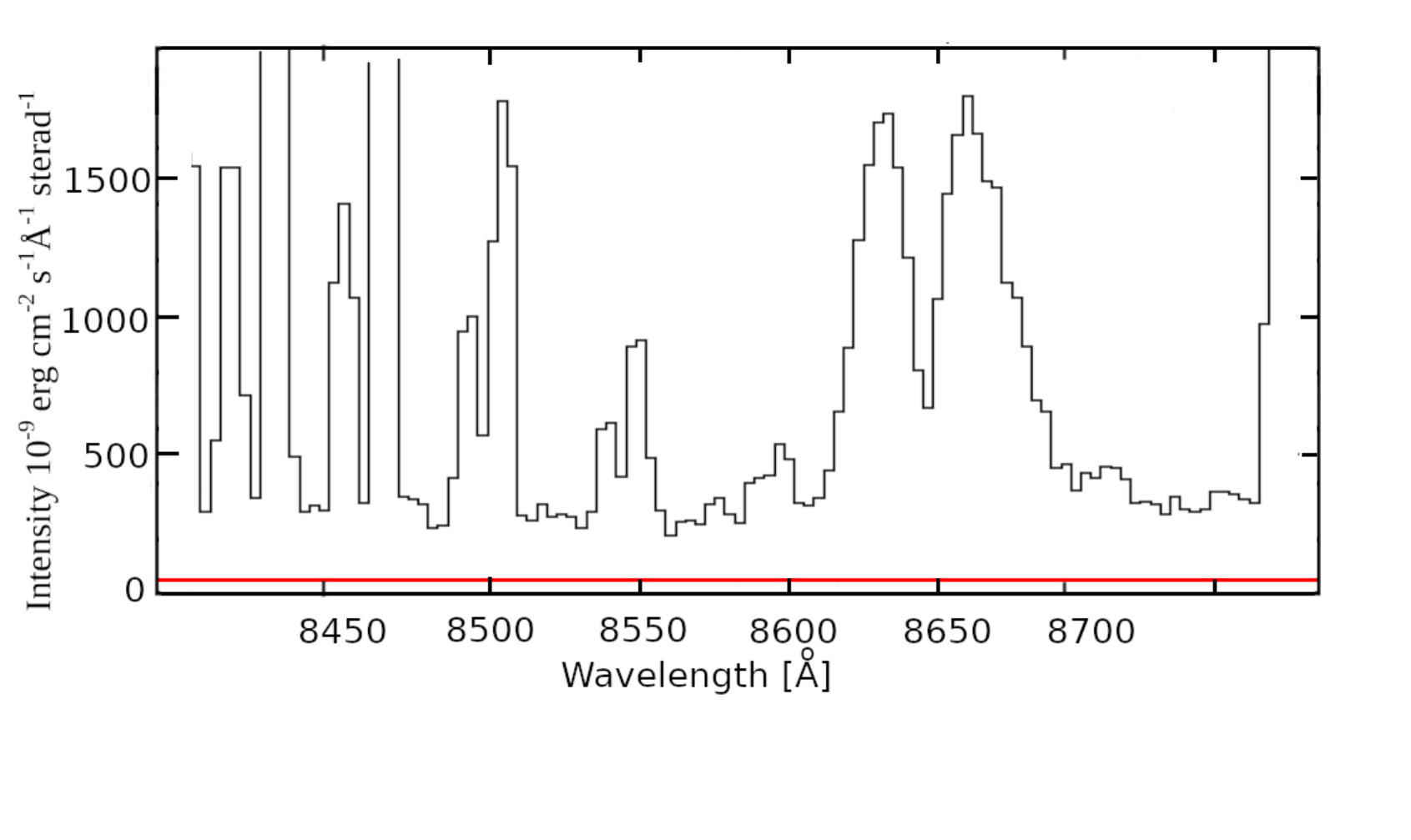}
\caption{
  Surface brightness in the A\_100\arcsec-off position in one
  observing block obtained on March 3, 2020, at an elevation of 37\degr.
  The level of expected zodiacal light surface brightness (approximately
  50\,cgs) is indicated with  a red line.}
\label{figure:agl}
\end{figure}

 A representative of our darkest total sky surface brightness spectra
 of the \caiitr\ wavelength region; the spectrum was obtained on March 3rd 2020
 at an  elevation of $37\degr$ (airmass 1.65), is shown in Fig. \ref
 {figure:agl}. The dominating component is the AGL. For comparison,
 the ZL+DGL contribution is $\le$50 cgs  \citep [see][]{OBrien2025}, that
 is $\le$20\% of the minimum sky brightness over this spectral range.

\subsection {Reduced {\em on--off} spectra}
\label {App:reduction}

\begin{figure} [h]
\includegraphics[width=88mm]{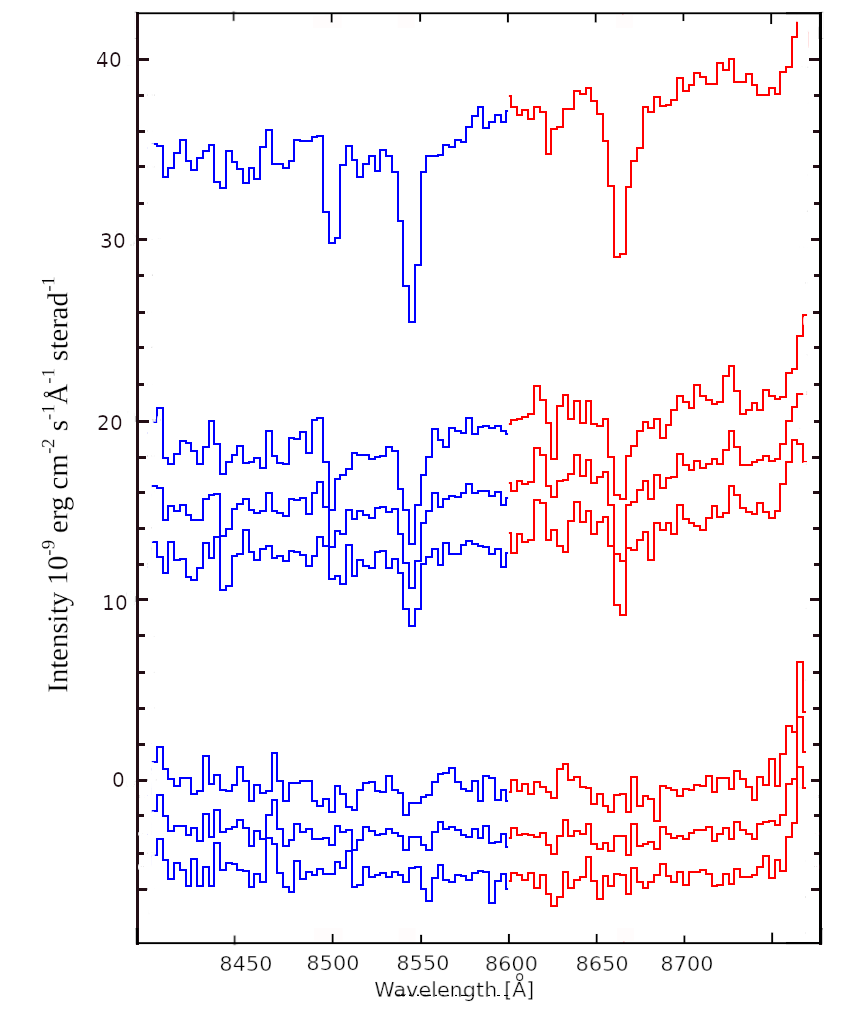}
\caption{ Final reduced spectra in the slit sections discussed in
  this paper: (from top to bottom) Bright\_rim, Core-up,
  Core-100$\arcsec$, Core-low, A\_low-off, A\_100\arcsec-off,
  A\_core-off.  The spectra for the sub-sets of 179 and 228 individual
  spectra, selected for minimum-AGL residuals, are shown as blue and
  red for $\lambda < 8600$\thinspace \AA\, and $\lambda >
  8600$\thinspace \AA\,, respectively.}
\label{figure:all}
\end{figure}

The reduced extracted {\em on--off} spectra in the slit sections
listed in Table \ref {Table:slitpos} are shown in Fig. \ref
{figure:all}. Blue and red are used to distinguish the spectra for the
sub-sets of 179 and 228 spectra, selected for minimum-AGL residuals,
respectively.

\section {The convolved GAIA RVS starlight  spectrum}
\label {App:convolution}

\begin{figure*} [h]
\centering
\includegraphics[width=160mm]{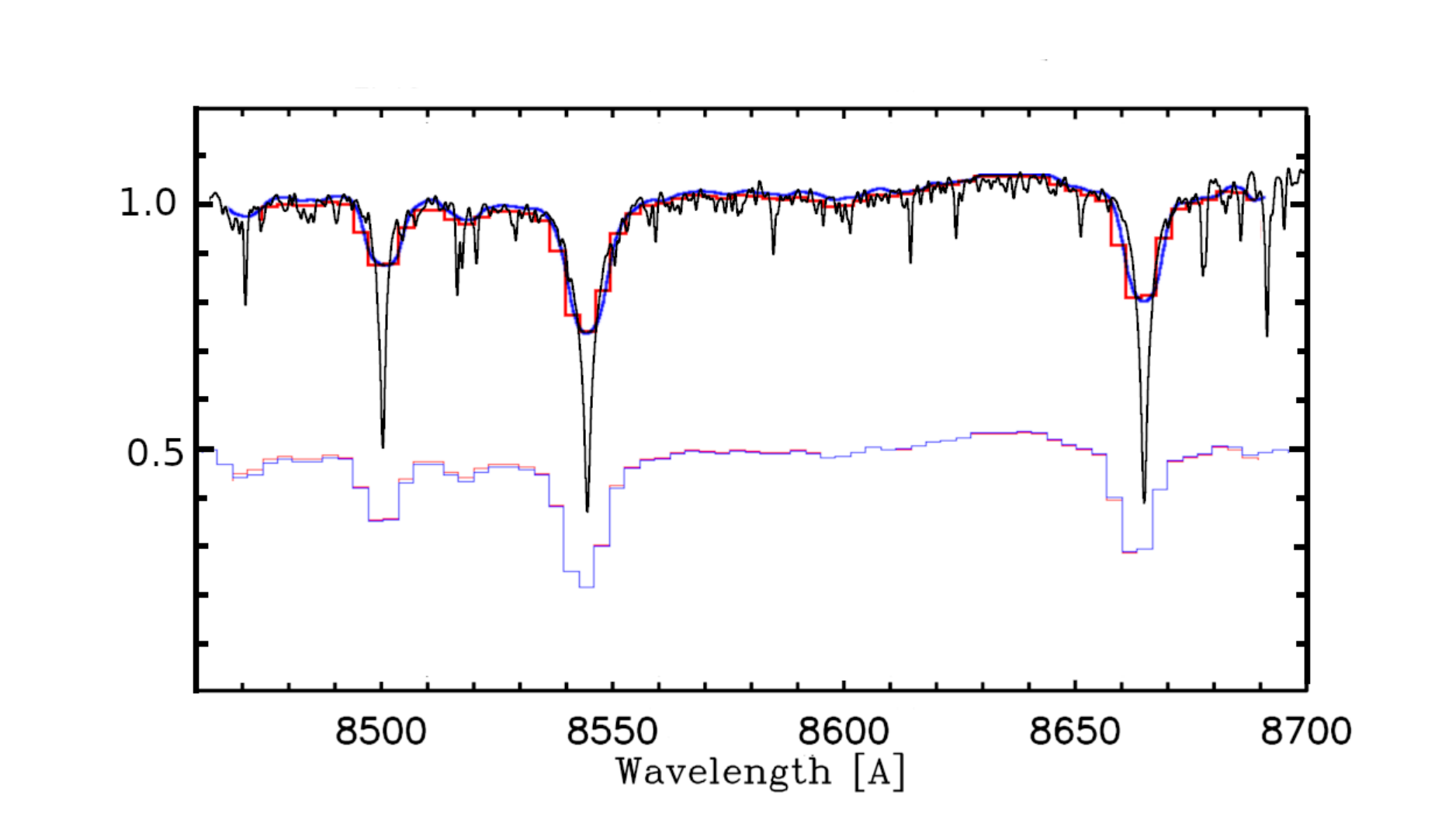}
\caption{Top: Summed normalised GAIA surface brightness assuming
  scattering function with asymmetry parameter g = 0.75 (Section \ref
  {App:shadowing}) (black), { convolved} using the FORS instrumental
  profile (blue), and resampled to the resolution of 3.28\thinspace
  \AA \ (red). Bottom: GAIA spectra smoothed using FORS instrumental
  profile and resampled to the resolution of 3.28\thinspace \AA
  \ assuming asymmetry parameter $g = 0.75$ (red), and 0.50 (blue)
  offset by -0.5 units.}
\label{figure:newgaia}
 \end{figure*}

  FORS instrumental profile (channel width 0.82\thinspace \AA \,),
  constructed using the FORS wavelength calibration lamp spectra, was
  resampled to correspond  to the GAIA spectrum channel width of
  0.1\thinspace \AA \,. This profile was used to { convolve}  the normalised
  GAIA RVS sum spectrum (Section \ref {sec:model_fit}) and then 
  resampled using SpecRess \citep{carnall2017} to the same channel
  width as the FORS spectra to be fitted. The original GAIA high
  resolution spectrum  for $g=0.75$, the { convolved} GAIA spectrum, and the GAIA
  spectrum resampled to the FORS binned 3.28\thinspace \AA\ channel
  width are shown in  the upper part of Fig. \ref{figure:newgaia} in black, blue, and
  red, respectively.  In the lower part this resampled GAIA spectrum together
with the resampled GAIA spectrum assuming $g=0.50$ are shown.

\section{EBL shadowing by the globule, attenuation factor $h(\tau)$}
\label {App:shadowing}

As has already been discussed in Section \ref {sec:components} the
globule not only attenuates the line-of-sight EBL but also scatters
the photons of the isotropic EBL into the observer's direction.  The
surface brightness difference, globule {\em(on)} minus transparent
surroundings {\em(off)}, is given by Eq. \ref{Eq1}:\\ $\Delta I_{\rm
  obs}(\lambda) = \Delta I_{\rm sca}(\lambda)-I_{\rm
  EBL}h(\tau)$\\ where we introduced the notation $h(\tau) =
1-e^{-\tau}-f_{\rm sca}(\tau)$, to be called the attenuation factor.

The surface brightness difference {\em on--off} is almost completely accounted to starlight
which has been scattered by dust in the globule.  This component, $\Delta I_{\rm sca}(\lambda)$,
when expressed as fraction of the infalling illumination by starlight over the sky, 
can be used to estimate  the scattered fraction, $f_{\rm sca}(\tau)$, of the EBL. 
In doing so we have to take into account that, while the EBL is isotropic, the distribution of 
the ISL is concentrated towards the Galactic plane. 
To estimate the infalling starlight at $\lambda \sim 8600$\thinspace \AA\,, 
effective for the illumination of the globule, we make use of the Pioneer all-sky map 
at the $R\_p$ band \citep{Gordon98} and the web page
\footnote{https://www.stsci.edu/\texttt{\char`\~}kgordon/pioneer\_ipp/Pioneer\_10\_11\_IPP.html}.
The average ISL surface brightness with equal weights
all over the sky is $\langle I_{\rm ISL}(R_p)\rangle$ = 151 \cgs; this estimate includes also the
 bright stars, $m<6.5$\,mag (Section 3.1.1 and Table 5, \citealt{mattilaetal2018}). It corresponds 
to the case where the dust would be isotropically scattering. However, the dust has a strongly
forward-directed scattering function as discussed in Section \ref {sec:model_fit} above. 
For a Henyey-Greenstein scattering function with an asymmetry parameter 
$g = \langle {\rm cos}\Theta \rangle = 0.75$ the weighted mean ISL sky brightness
as seen by the globule is  $\langle I_{\rm ISL}(R_p)\rangle$ = 253 \cgs\,.

In order to {refer these  $R_p$ band values, with a pivot wavelength of 6441\thinspace \AA\,, 
to $\lambda$ = 8600\thinspace \AA\,}
we make use of the \citet{lehtinen_mattila_2013} Milky Way SED model. It gives
{ the scaling factor} $I_{\rm ISL}(8600)/I_{\rm ISL}(R_p)$ = 0.88. Then, the effective sky brightness
weighted according to $g = 0.75$, is  $\langle I_{\rm ISL}(8600)\rangle$ = 223 \cgs\,.
If weighted according to  $g = 0.50$, the effective sky brightness would be  
$\langle I_{\rm ISL}(8600)\rangle$ = 192 \cgs\,.
 
The optical depth through the core of \DCL\ is > 10; thus, the transmission $e^{-\tau}$ 
is negligible. For the bright rim the optical depth can be estimated via two methods: first, via 
direct $JHK_s$ photometry \citep{kainulainenetal2007} and secondly, via the observation that the maximum 
surface brightness of scattered light in globules and dark nebula cores occurs at an optical depth 
of $\sim 1.5 - 2.5$ \citep{witt1974, mattilaetal2018}. We adopt $\tau(8600) = 2$.

With reference to the results for line 2 (8542 \thinspace \AA\,) in Tables \ref {Table:results_GAIA} and  \ref {Table:results_Bright} we adopt for $h(\tau) I_{\rm EBL}$  
the values 1.5 and 1.1 \cgs\, for the Core-up and Bright\_rim, respectively.
The $\Delta I_{\rm sca}(8600)$ values are then { estimated to be 14.5 -- 20.8 and 37.7 \cgs\ for the
 the three core sections and the Bright\_rim, respectively. These values result in the ratios
 $\Delta I_{\rm sca}/\langle I_{\rm ISL} \rangle$=$f_{\rm sca}(\tau)$ = 0.065 -- 0.093 and 0.17.
for the  core sections and the Bright\_rim, respectively.}

With these estimates we end up with {value  $h(\tau) = 0.92$ as the best estimate to be used 
for the three core positions; it is the mean of the values for Core-up and Core-low and it coincides 
with the value for Core-100\arcsec\,. The value $h(\tau) = 0.69$
is obtained for the Bright\_rim} (see Table \ref{Table:attfact}). 
With an estimated error of $\pm10\%$ for both 
$\Delta I_{\rm sca}$ and $\langle I_{\rm ISL}\rangle$ we estimate the systematic error to be 
$\pm2\%$ and $\pm4\%$ for the Core-up and Bright\_rim, respectively. However, for the case when
the bright rim is used as a template to fit the dark core spectrum (see Section \ref{sec:model_fit2}), 
{$h_{\rm eff}=0.58$} with an error of $\pm2\%$, as well. If $g=0.50$ had been adopted
insted of  $g=0.75$ the attenuation factors of 0.91, 0.66 and 0.58 would have resulted, instead.
The differences relative to the  case  $g=0.75$ are small and cause no changes to our
EBL values or their error estimates.

\begin{table*}
\begin{minipage}{100mm}
\caption{ Empirical determination of 
{the attenuation factor at  $h(\tau) = 1-e^{-\tau}-f_{\rm sca}(\tau)$}  at 8600\thinspace \AA\, 
using scattered starlight from \DCL. }

\begin{tabular}{lrrrr}
\hline                          
\hline                          

\label {Table:attfact}          & Core-up& Core-low & Core-100\arcsec\, & Bright\_rim \\
                                & [cgs]  & [cgs]    & [cgs]             &[cgs]      \\
\hline
$\Delta I_{\rm obs}$             & 19.3   &13.0     & 16.5              &36.6      \\
$I_{\rm EBL}h(\tau)$\tablefootmark{a,b} & 1.5 & 1.5 & 1.5 & 1.1 \\
$\Delta I_{\rm sca}$             & 20.8 & 14.5 & 18.0  & 37.7\\
 $\langle I_{\rm ISL}\rangle$ \tablefootmark{c}   &223   &223 &223 & 223     \\
$\Delta I_{\rm sca}/\langle I_{\rm ISL} \rangle$=$f_{\rm sca}(\tau)$ & 0.093 & 0.065 & 0.081& 0.17 \\
$\tau(8600)$                        & $> 10$ &  $> 10$ & $> 10$ & $\sim$2    \\
$e^{-\tau}$                            & 0  & 0 & 0  &  $\sim$14     \\
 $h(\tau)$                      &  {0.907}  &0.935 & 0.919&  0.69 \\
\hline                          
\end{tabular}
\tablefoottext{a}{$I_{\rm EBL}h({\tau})$ is used here as a correction term to be added to 
$\Delta I_{\rm obs}$; different values of $h({\tau})$ have been iteratively used. \newline}
\tablefoottext{b}{The value adopted for $I_{\rm EBL}$ is 1.62 \cgs. \newline}
\tablefoottext{c}{The effective sky brightness at $\lambda$ = 8600\thinspace \AA\,.}
\end{minipage}
\end{table*}

\section{Aperture correction}
\label{App:Cal_aperture}
The intensity (surface brightness) of an extended source in \cgs\, is given by
\begin{equation}
I(\lambda) = \frac{S(\lambda) T(A)}{\Omega} C(\lambda),
\end{equation}
where  $C(\lambda)$ is the signal in instrumental 
units (ADU), $\Omega$ the solid angle of the 
aperture in steradians, and $S(\lambda)$ the sensitivity function  
in units of $10^{-9}$~erg\,cm$^{-2}$s$^{-1}$\AA$^{-1}$/ADU, as determined from 
the standard-star observations. $T(A)$ is the aperture correction, which is
required when calibrating an extended surface brightness with standard star observations.   
In the measurement of a uniform extended source  
the flux that is lost from the solid angle $\Omega$, as defined by the focal plane aperture, 
is compensated by the flux that is scattered and diffracted into the aperture from the sky
outside of the solid angle $\Omega$; thus $T(A) = 1$. However, for the standard stars
used to calibrate the surface brightness in physical flux units the aperture correction, 
to be noted by $T(A)^*$, is < 1. Only the fraction  $T(A)^*$ of the flux from the star is contained 
within the aperture, while the fraction $1 - T(A)^*$ is lost outside the aperture  
by diffraction and scattering in the telescope and instrument optics.

\subsection{Aperture correction for standard stars}
Spectrophotometric standard stars were observed in the ESO calibration service program 
using a {5\arcsec} slit. The spectra were extracted from the sum of stellar flux in the extraction 
window of 3.0\arcsec\ width in the spatial direction. In order to estimate the flux fraction 
falling outside of the extraction window we used the stacked spectrum of nine standard star 
observations available for our program. The flux fraction up to a distance of 20\arcsec \ was found 
to be 0.172 with an estimated error of $\pm0.02$  The wavelength range covered was 8300 - 8800\thinspace \AA\,.
This estimate can be compared with the results presented in \citet{mattila_lehtinen_etal_2017}, sect. 6.4. 
They also used FORS2 at UT1 and a similar spectrometer set-up. The wavelength range covered
three slots, 3500--4250, 4250--5000, and 5000--6000\thinspace \AA\,. From a stacked spectrum of 17 standard 
stars they estimated the energy falling outside of the $2.5\arcsec \times 5\arcsec$ 
extraction slot, up to a distance of 20\arcsec.  No wavelength dependence was found, 
and the mean value for these three slots was $1 - T(A)^*=0.143\pm0.006$, which is within the errors 
consistent with our present result for 8300 - 8800\thinspace  \AA\, 

The fraction of energy falling outside of 20\arcsec\  was estimated by \citet{mattila_lehtinen_etal_2017} 
using measurements (FORS1 at UT2) of Sirius' aureole:
 $1 - T(A)^*$ was found to be 0.029 for the flux fraction between 20\arcsec -- 100\arcsec,  
and 0.038 between 100\arcsec -- 2\degr\,; the values were closely the same in blue and visual.
With an estimated uncertainty of $\pm50$\% for each of these estimates the total error 
was estimated to be $\sim \pm 0.05$. 
In spite of the wavelength difference we adopt this value for our present wavelength slot as well.

Summing up, at 8300 - 8800\thinspace  \AA\, the total fraction of the standard stars' flux  lost between 3\arcsec \ 
and  2\degr\, is estimated to be $1-T(A)^* = 0.24\pm 0.06$ and thus, the aperture correction is 
$T(A)^* = 0.76 \pm 0.06$. 
Because of the aperture correction the standard star fluxes to be used for the calibration of the
surface brightnesses are, instead of their list values $I_{\lambda}$, given by $T(A)^*I_{\lambda}$.
As a consequence, the surface brightness measurements, scaled by the standard star list values, 
have to be re-scaled down by the factor $T(A)^*$.

\section{Synthetic ISL spectrum model for 8450 - 8700\thinspace  \AA\, }
\label{App:ISLmodel}

\begin{figure*} [h]
\sidecaption
  \includegraphics[width=12cm]{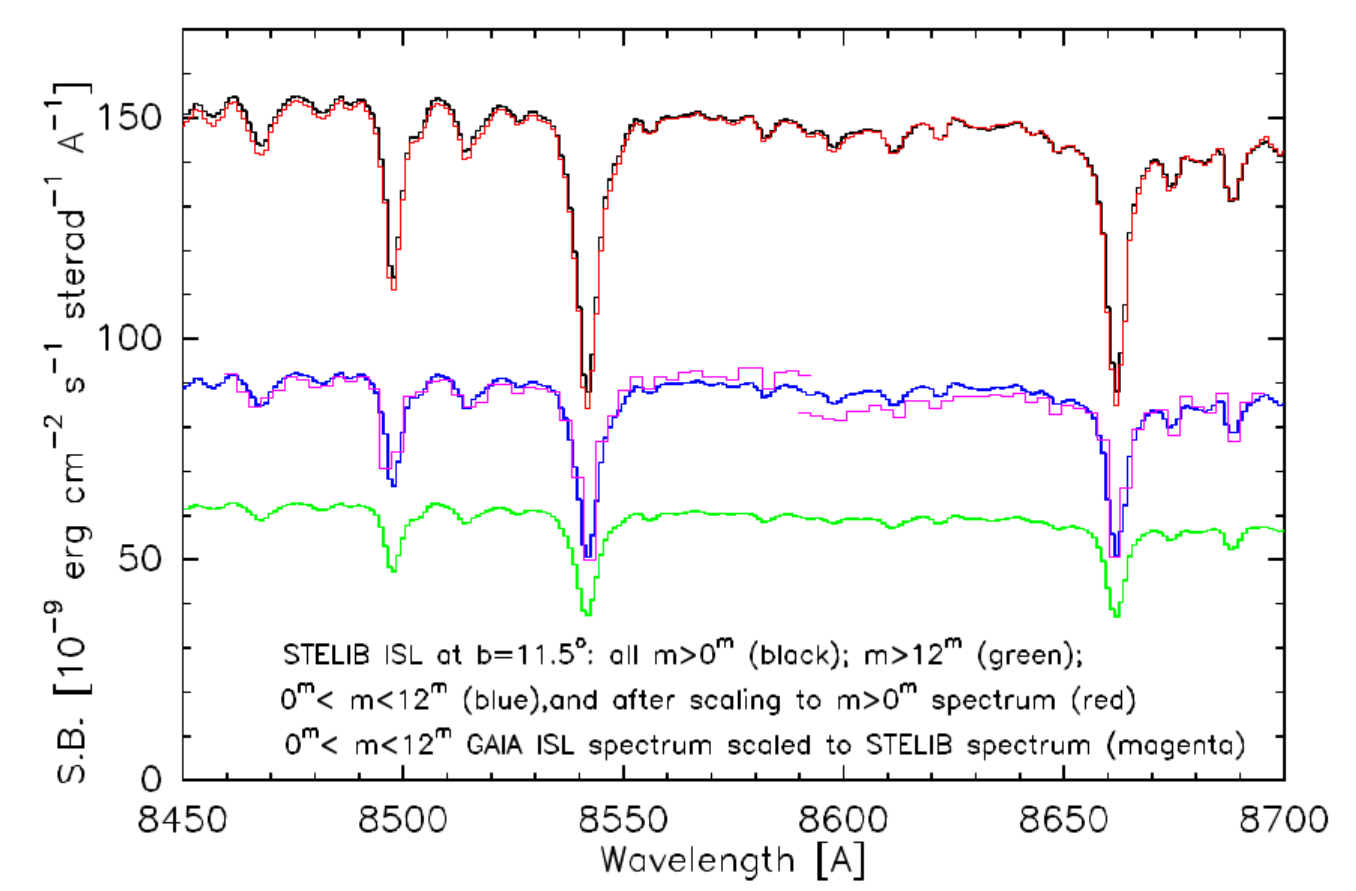}
\caption{Synthetic ISL spectra based on the STELIB spectral library. The spectra
are shown for different magnitude intervals: total starlight $m>0$ mag (black);
$m>12$ mag (blue); $0<m<12$ mag (green). In order to demonstrate the agreement
between the total ISL and the $0<m<12$ mag spectra, a scaled version of the
latter (in red) is overplotted on the total spectrum. A suitably re-scaled  GAIA/RVS
spectrum is shown as the magenta line overplotted upon the model spectrum for
 $0<m<12$ mag (blue line). Because of the difference in baseline slopes, the GAIA/RVS
spectrum has been cut at 8590\thinspace \AA\, and is shown as two pieces.}
\label{STELIB_models}
\end{figure*}

As has been described in Section \ref {sec:model_fit} a major part of the ISL intensity
at 8500 - 8700\thinspace  \AA\, is covered by the sum of RVS spectra available
in the GAIA DR3 for stars brighter than 12\,mag. In this appendix we shall 
investigate to what extent the contribution by the fainter stars with $G_{RVS} > 12$\,mag  
influences the \caiitr\  spectrum at 8500 - 8700\thinspace  \AA\,. 

For modelling of the ISL spectrum we make use of the spectral synthesis method which 
has commonly been used to analyse the contributions by different stellar populations 
to the total light of an external galaxy or star cluster (see e.g. \citealt{bruzual03}). 
Here, we have addressed the opposite problem:  given the number densities and  
the spatial distributions of the different spectral types 
of stars in the Solar neighbourhood, what is the spectrum of their integrated light
(ISL) for different magnitude intervals and in different directions over the sky. 
Our results for the ISL spectrum at 8450 - 8700\thinspace \AA\, make use of the 
spectral synthesis modelling as presented in \citet{mattila_vaisanen_etal_2017}. 

There, a simple model of the  Galactic structure was adopted 
in which stars and dust are distributed in plane parallel layers.
Stars were divided into 72 spectral groups covering the different parts 
of the HR-diagram. The division was made according to the 
approach of \citet{flynn06} based on their analysis of the {\it Hipparcos}
database. The spectral groups were compiled under the following 
seven categories: {\it (i, ii)} main sequence (thin and thick disk), {\it (iii, iv)} clump 
stars (thin and thick disk),  {\it (v, vi)} old giants (thin and thick disk), 
{\it (vii)} young giants. Each group was characterised by the following parameters:
the mean absolute magnitude $M_V$; the number density $D(0)$ and the stellar 
emission coefficient $j_i(0)$ in the Galactic plane, $z=0$; and the scale 
height $h_z$ for a distribution of the form $D(z) = D(0)$sech$(z/h_z)$. Because of the 
limited distance range of {\it Hipparcos}, its coverage for the supergiants was sparse. 
This group was, therefore, complemented by using the compilation of \citet{Wainscoat92}.

For the synthetic model of the ISL spectrum a spectral library is needed with good 
coverage of spectral types and luminosity classes as well as a sufficient spectral resolution, 
corresponding to or better than that of  the observed spectra. The STELIB library 
\citep{LeBorgne03}, with its
spectral resolution of $\la 3$\thinspace \AA\,(FWHM), matches well the resolution requirement 
for our VLT/FORS2 spectra. From this library the best template stars for each
stellar group were chosen: besides the spectral class and absolute magnitude 
$M_V$, also the colour indices $B-V$ and $V-I_c$ were used as selection criteria.

ISL model spectra covering the \caiitr\  wavelength region are shown in 
Fig. \ref{STELIB_models}. The spectra are  for the galactic 
latitude $|b|=11.5\degr$. No galactic longitude dependence is included in the model, 
and symmetry is assumed w.r.t. the galactic plane. The adopted galactic latitude
corresponds to a weighted mean of contributions from Milky Way areas
around the direction of the \DCL\,. The spectra are shown for three different magnitude
intervals, $m>0$ (black), $0<m \le 12$ mag (blue) and $m>12$ mag (green). The relative
intensity levels are seen to correspond to the starcount results, presented in Section \ref {sec:model_fit}: 
approximately 40\% of the total ISL is contributed by stars with 
$m > 12$ mag. In order to see whether the missing stars have a strong
effect on the \caiitr\  lines we have overplotted the spectrum for $0<m<12$\,mag,
suitably scaled (red line), upon the total ISL spectrum. The good agreement between 
the black and red lines demonstrates that the spectrum for $0<m\le12$ mag
 is a good representative also for the total ISL.
  
Being based on the STELIB  library, the model spectra shown in Fig. \ref{STELIB_models} 
have a resolution of $\la 3$\thinspace \AA\,(FWHM) and have been plotted with a step size
of 1\thinspace \AA\, corresponding to the original database. In order to enable
comparison with STELIB, we have smoothed the GAIA RVS spectrum to 
a resolution of 2.7\thinspace  \AA\,. Suitably re-scaled, it has been
overplotted in  Fig. \ref{STELIB_models} as the magenta line on the model spectrum for
 $0<m\le12$\,mag (blue line). A good agreement is demonstrated between the GAIA RVS and
the STELIB synthetic spectra.
 
\end{appendix}
\end{document}